\documentclass[aps,pra,showkeys,superscriptaddress,amsmath,floatfix,twocolumn,twoside,amssymb,reprint,notitlepage,nofootinbib]{revtex4}

\usepackage{epsfig}
\usepackage{graphicx}
\usepackage{bm}
\usepackage{dcolumn}
\usepackage{epstopdf}
\usepackage{lipsum}
\usepackage{blindtext}
\usepackage{natbib}
\usepackage{hyperref}
\usepackage[utf8]{inputenc}

\graphicspath{{./}{figures/}}

\newcommand{\be}{\begin{equation}}
\newcommand{\ee}{\end{equation}}
\begin{document}
	
    \title{A minimal model for structure, dynamics, and tension of monolayered cell colonies}

\author{Debarati Sarkar}
\email{d.sarkar@fz-juelich.de}
\author{Gerhard Gompper} 
	\email{g.gompper@fz-juelich.de}
\author{Jens Elgeti}
\email{j.elgeti@fz-juelich.de}

\affiliation{Theoretical Physics of Living Matter, 
Institute of Biological Information Processing and Institute for Advanced Simulation, 
Forschungszentrum J\"ulich, D-52425 J\"ulich, Germany}

\begin{abstract}
The motion of cells in tissues is an ubiquitous phenomenon. In particular, in monolayered cell colonies in vitro, pronounced 
collective behavior with swirl-like motion has been observed deep within a cell colony, while at the same time, the colony remains 
cohesive, with not a single cell escaping at the edge.
Thus, the colony displays liquid-like properties inside, in coexistence with a cell-free ``vacuum" outside.   
How can adhesion be strong enough to keep cells together, while at the same time not jam 
the system in a glassy state? What kind of minimal model can describe such a behavior? Which other signatures of activity arise
from the internal fluidity? 
We propose a novel active Brownian particle model with attraction, in which the interaction potential has a broad minimum 
to give particles enough wiggling space to be collectively in the fluid state.
We demonstrate that for moderate propulsion, this model can generate the fluid-vacuum coexistence described above. 
In addition, the combination of the fluid nature of the colony  with cohesion leads to preferred orientation of the 
cell polarity, pointing outward, at the edge, which in turn gives rise to a tensile stress in the colony -- as observed 
experimentally for epithelial sheets. For stronger propulsion, collective detachment of cell clusters is predicted.
Further addition of an alignment preference of cell polarity and velocity direction results in enhanced coordinated, 
swirl-like motion, increased tensile stress and cell-cluster detachment.  
\end{abstract}

\pacs{Valid PACS appear here}

\maketitle

\section{Introduction}
\textbf{M}any fundamental biological processes, like embryogenesis, wound healing or cancer/tumor invasion require 
cells to move collectively within tissues \cite{martin2004parallels,friedl2004collective,lecaudey2006organizing}. 
The physics underlying these processes ranges from understanding actin polymerization and tread-milling for force
generation \cite{wodnicka-1996,sheetz-2007} and single cell migration \cite{horwitz-1996,theriot-nat}, to the collective 
behavior of many migrating cells \cite{nishida-04,puliafito-pnas,mcgrath-07}. Here, we focus on an  
observation from monolayers of migrating Madin-Darby canine kidney (MDCK) cells on surfaces, a prototypical model system 
for collective cell migration. Experimental observations reveal 
large-scale collective motion, like swirls, within the bulk of young monolayers \cite{angelini-10, angelini-11}, 
thus the display of fluid-like properties,  before jamming occurs as the epithelial sheet matures 
\cite{manning-natp-2015,manning-prx-2016,garcia-2015,prost-2018}. Interestingly, as an initial 
colony expands, no cells detach from the boundary --- even though the bulk of the tissue remains clearly 
fluid-like \cite{puliafito-pnas}. 
Cohesion is strong enough that fingers of many cells can protrude at the propagating tissue front 
without cell detachment. Even stronger-pulling ``leader cells'' do not detach 
\cite{bonder-pnas-2003,silberzan-pnas-2007,silberzan-bpj,spatz-2018}. Cells are thus in a 'liquid-vacuum' 
\footnote{We use the term 'vacuum' here somewhat loosely to represent a phase of extremely low cell density} 
coexistence regime. Even more surprising, pioneering experiments have revealed that these expanding colonies are under 
tensile stress \cite{trepat-nat,silberzan-cellbio-2014}. This raises the question how this liquid-vacuum coexistence, in 
combination with strong collective motion and tensile stress, can be captured and understood from a minimal 
physical model.

An active Brownian particle (ABP) model \cite{marchetti-PRL,hagan-PRL,wysocki_EPL2014,Elgeti_RoPP2015} for cells with 
standard attractive Lennard-Jones (LJ) interactions has been proposed 
to study cell colonies \cite{redner-2013,fielding-2015,filion-2016}. However, only solid-vacuum (no fluidity of the 
condensed phase), or liquid-gas (finite cell density in the dilute phase) coexistence has been obtained. 
The coexistence of liquids with a very-low-density gas phase is of course well known in many equilibrium systems. 
In the biological context, for example, lipid-bilayer membranes are liquid in nature, but the critical-micelle concentration 
is very low, so that lipids essentially never detach from the membrane. 
In the modelling and simulation of lipid membranes, a similar
problem of attractive interactions and fluidity exists as for cell monolayers -- too strong attraction leads to solidification.
In coarse-grained simulations of lipid bilayers, this problem was overcome by an interaction potential with an extended range 
compared with the standard Lennard-Jones potential,  which provides strong adhesion while 
still giving enough wiggle room for the molecules that the membrane to remains the fluid phase \cite{Cooke_PRE2005}. 
In the spirit of minimal modeling, we propose an active Brownian particle (ABP) model for the cells, combined with a similar 
longer-range interaction potential as employed successfully for the membrane lipids. 
We demonstrate that the LJ potential with a 
wider attractive basin indeed opens up a region in phase space that displays liquid-vacuum 
coexistence. The size of the liquid-vacuum region expands as the basin width of LJ potential increases.
The fluidity of the condensed phase implies the emergence of several interesting behaviors,
like a tensile stress within the colony due to a preferred orientation of the boundary cells to the 
outside, as awell as the detachment of cell cluster above a size threshold. 
When a coupling of cell polarity to the instantaneous direction of motion -- which is significantly affected by the interactions 
with the neighboring cells --  is introduced, the formation of swirls and collective cell detachment is strongly
enhanced.  


\section{Results and Discussion}
\label{sec:res}

\subsection{Active Brownian particles with attraction}
\label{sec:macro}

Liquid-vacuum coexistence requires strong inter-particle adhesion, so that cells can not detach from the main colony.  
Concurrently, the adhesion has to provide enough wiggling room that the cells remain locally mobile inside the 
condensed phase and provide fluidity to the colony.  A long-range coordinated motion of cells, like fingering or swirls, 
then already emerges to some extent from the self-organized motion of cells which all vary their propulsion direction
independently and diffusively. However, pronounced correlations are 
found to require additional alignment interactions of cell orientation and direction of motions. 
Here, the effect of neighbors pushing (or pulling) a cell in a certain direction is assumed to
induce a reaction in the cell to reorient and align its propulsion direction with its instantaneous
velocity direction.
 
The ABP model, where each particle is a sphere (in 3D) or a disc (in 2D) which undergoes rotational Brownian motion and 
additionally experiences a body-fixed driving force of constant magnitude, was developed to describe active motion 
on the microscale \cite{garcia-2015,vincent-pnas}.
This model displays a rich phase behavior, most notably motility-induced phase separation 
\cite{marchetti-PRL,hagan-PRL,Burttinoni_PRL2013,wysocki_EPL2014}, where persistence of motion and short-range repulsion 
induce cluster formation.  Addition of a Lennard-Jones attraction leads to the formation of 
arrested clusters for small propulsion \cite{redner-2013,mognetti-PRL-2013,fielding-2015}.
In order to obtain liquid-like properties at strong adhesion, we follow the spirit of Ref.~\cite{Cooke_PRE2005} and 
propose an interaction potential with an extended basin of width $\bar\sigma$, so that 
\[V_{m}= \left\{
\begin{array}{lr}  
4\epsilon \left[(\sigma/r)^{12}-(\sigma/r)^{6}\right], & 0 < r\leq 2^{1/6}\sigma \\
-\epsilon, &   2^{1/6} \sigma < r \leq \tilde{r} \\
4\epsilon \left[ (\sigma/(r-\bar{\sigma}))^{12}-(\sigma/(r-\bar{\sigma}))^{6} \right], &  \tilde{r} < r\leq r_{cut} 
\end{array}
\right.
\]
(see Appendix A Fig.~6).  
Here, $\sigma$ is the particle diameter, $\tilde{r} =(2^{1/6}\sigma+\bar{\sigma})$, and $\epsilon$ is the 
interaction strength. 
This modified interaction provides a short-range repulsion or volume exclusion for the particles with 
separation $r<2^{1/6} \sigma$, a force-free regime for $2^{1/6} \sigma < r \leq \tilde{r}$, and a  
long-range attraction for particle separation, $\tilde{r}<r< r_{cut}=2.5\sigma$. 

For activity, each ABP is subjected to a constant active force $f_0$ along a body-fixed propulsion direction 
${\bf \hat{n}}_i=(\cos\theta_i, \sin\theta_i)$. The orientation $\theta$ undergoes diffusive reorientation, and may 
additionally experience alignment forces. Time evolution follows a Langevin dynamics,  
\be\label{eq:1}
\begin{split}
m\ddot{\bf r}_i &=-\gamma \dot{\bf r}_i+ {\bf F}_{i}({\bf r}_i) + f_0 {\bf \hat{n}}_i+ \sqrt{2D} {\bf \eta}_i^{T},\\
                     \dot{\theta}_i &= \sqrt{2 D_r }\eta_i^{R}. 
\end{split}
\ee
Here, ${\bf F}_{i}=-\nabla_i V$ describes the interaction with other cells with the total potential $V$ as a sum 
of all pair interactions, and $f_0=v_0\gamma$ is the driving force which results in a self-propulsion velocity $v_0$ 
for an isolated cell experiencing a drag force due to substrate friction with drag coefficient $\gamma$, which
is related to the thermal translational diffusion coefficient $D=k_BT/\gamma$ by the Einstein relation.
Similarly, $D_r$ is the rotational diffusion coefficient. 
The noise forces $\eta$ are assumed to be Gaussian white-noise variables with $\langle\eta_i(t)=0\rangle$ 
and $\langle \eta_i(t) \eta_j(t^{\prime}) \rangle=\delta_{ij}\delta(t-t^{\prime})$.  
However, note that this is an active system, and thus both diffusion processes can in principle be 
independent active processes with different amplitudes, and thus do not need to satisfy the Einstein 
relation or fluctuation-dissipation theorem. 
In order to emphasize the importance of rotational over translational diffusion, 
we choose $D_r \sigma^2/D=3$.

The cohesive nature of modeled cell colonies depends on the competition between adhesion and self-propulsion. 
A key parameter is the potential width $\bar{\sigma}$, which controls the fluid-like consistency of the colony.
Further details about model parameters can be found in the supporting information.
In the simulation results described below, all quantities are reported in dimensionless units based on thermal energy $k_B T$, 
particle diameter $\sigma$, and rotational diffusion time $\tau_r=1/D_r$.  We characterize the system by three 
dimensionless numbers, the P{\'e}clet number, $Pe=v_0 \sigma^2/D = 3 v_0 \tau_r /\sigma$ which quantifies the activity of the system,
the adhesion strength $U=\epsilon/(k_B T)$, and the potential width $\bar\sigma/\sigma$ which determines 
the wiggle room of the cells.

In order to introduce local velocity-orientation alignment, we assume alignment between propulsion direction and  
velocity for each cell individually \cite{szabo-06,basan-2013,dauchot-2015}.
In our simple stochastic model, Eq.~\ref{eq:1}, the orientation dynamics in this case is determined by 
\be\label{eq:2}
\dot{\theta}_i = \sqrt{2 D_r }\eta_i^{R} -k_e D_r \frac{\partial}{\partial \theta} ({\bf{n}}_i.{\bf{v}}_i)
\ee   
Here, $k_e$ is the strength of particle alignment.
The alignment force can be interpreted as arising from a pseudo potential $V_a=-(k_e/2)({\bf v} \cdot {\bf n})$,
acting only on the orientation $\bf n$, but not on $\bf v$. Unless noted otherwise, results concern systems 
withouth velocity-alignment interactions (i.e. $k_e=0$). 


\subsection{Liquid-Vacuum Coexistence}

\begin{figure*}[ht]
\centering 	
\includegraphics[width=0.70\textwidth]{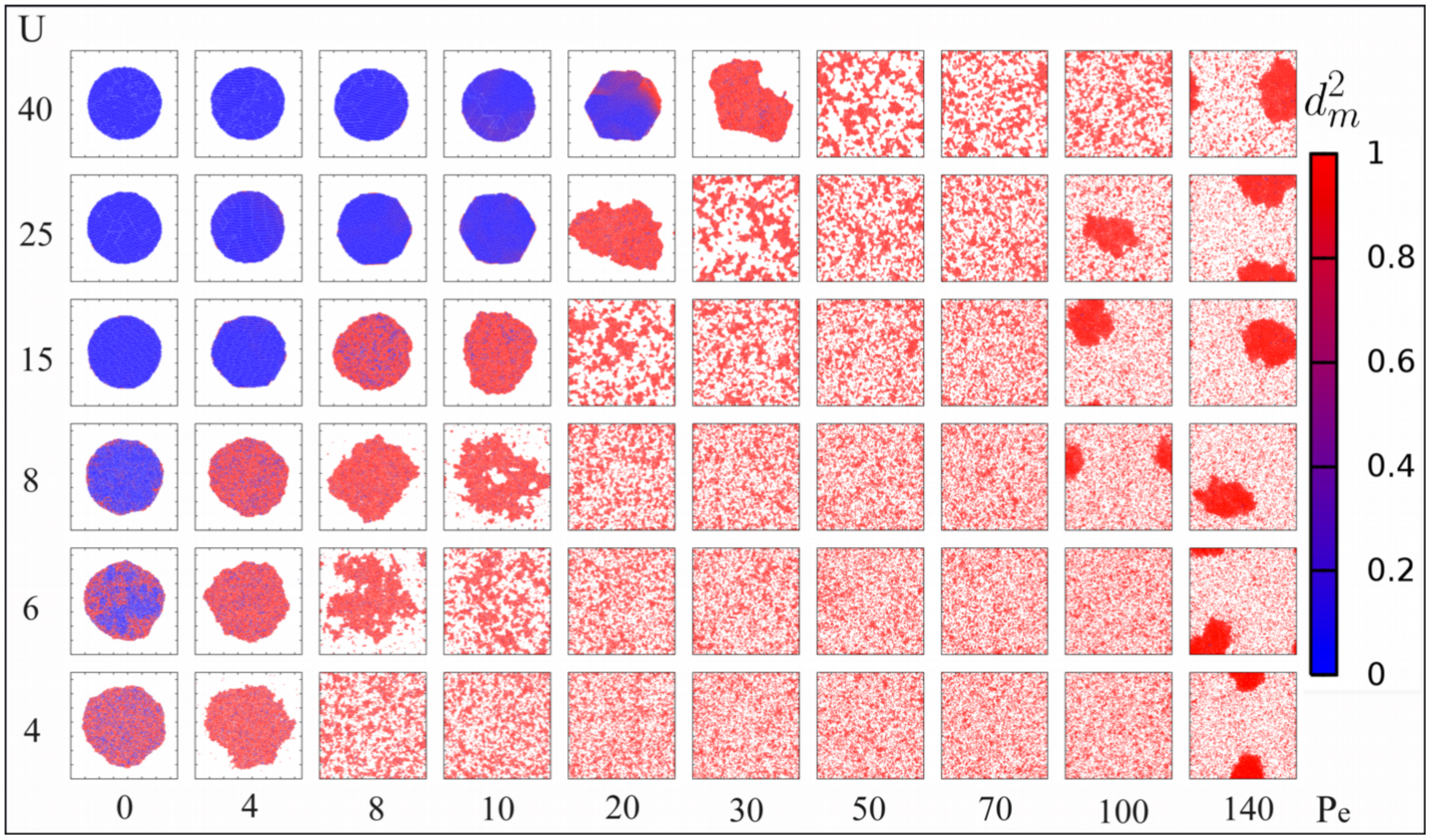}  
\caption{Phase diagram illustrated by snapshots at the end of simulation. Here we plot the mobility profile as a function 
of attractive interaction $U$ and activity $P_e$. We start the simulation with initial circular patch. The colour code 
defines the magnitude of mobility. Blue means immobile and red means highly mobile colony. 
($\bar{\sigma}/\sigma = 0.3$; overall packing fraction is $0.274$). See SI Movies S1 and S2 for the formation and dynamics 
of the cohesive colony at liquid vacuum coexistence for $U=40$, $Pe=30$.}
\label{fig:den}
\end{figure*}

We begin our analysis by exploring the available phase space spanned by activity $Pe$ and adhesive strength $U$.
The system is initialized with a circular cell colony with $N=7851$ particles and a diameter $100 \sigma$ 
(packing fraction $\phi=0.79$) in a square simulation box of linear size $150\sigma$.  The resulting phase behavior 
as a function of $Pe$ and $U$ is displayed in Fig.~\ref{fig:den}. Here, snapshots of particle conformation
after long simulation time ($t=3300 \tau_r$), together with particle mobility, measured by the mean squared displacement 
\be\label{eq:3}   
d_m^2=({\bf r}_i(t+t^{\prime})-{\bf r}_i(t))^2
\ee
averaged over several reorientation times $t^{\prime}=12\tau_r$, are employed to characterize the phases. 
Figure~\ref{fig:den} shows that a line $Pe \simeq U$ separates a homogeneous gas phase at $Pe>U$ from a two-phase coexistence
regime for $Pe<U$. For low activity ($Pe\ll U$), the condensed phase is solid, where particles do not show any 
significant movement, i.e. $d_m^2\approx0$. As activity increases and approaches $Pe\simeq U$, cells become mobile ($d_m^2>1$). 
Finally, for large $Pe \gtrsim 100 \gg U$, attraction becomes negligible and conventional motility-induced phase separation is observed. 
A simple calculation, which equates the propulsion force with the maximum of the attraction force, reveals that the 
detachment of particle pairs occurs at  $Pe= 2.4 U$; for larger $Pe$ the adhesive force is no longer strong enough to 
keep particles together.  Note that thermal fluctuations are usually rather small in this study (because $U \gg 1$).

\begin{figure*}[!htb]
\centering
\includegraphics[width=0.48\textwidth]{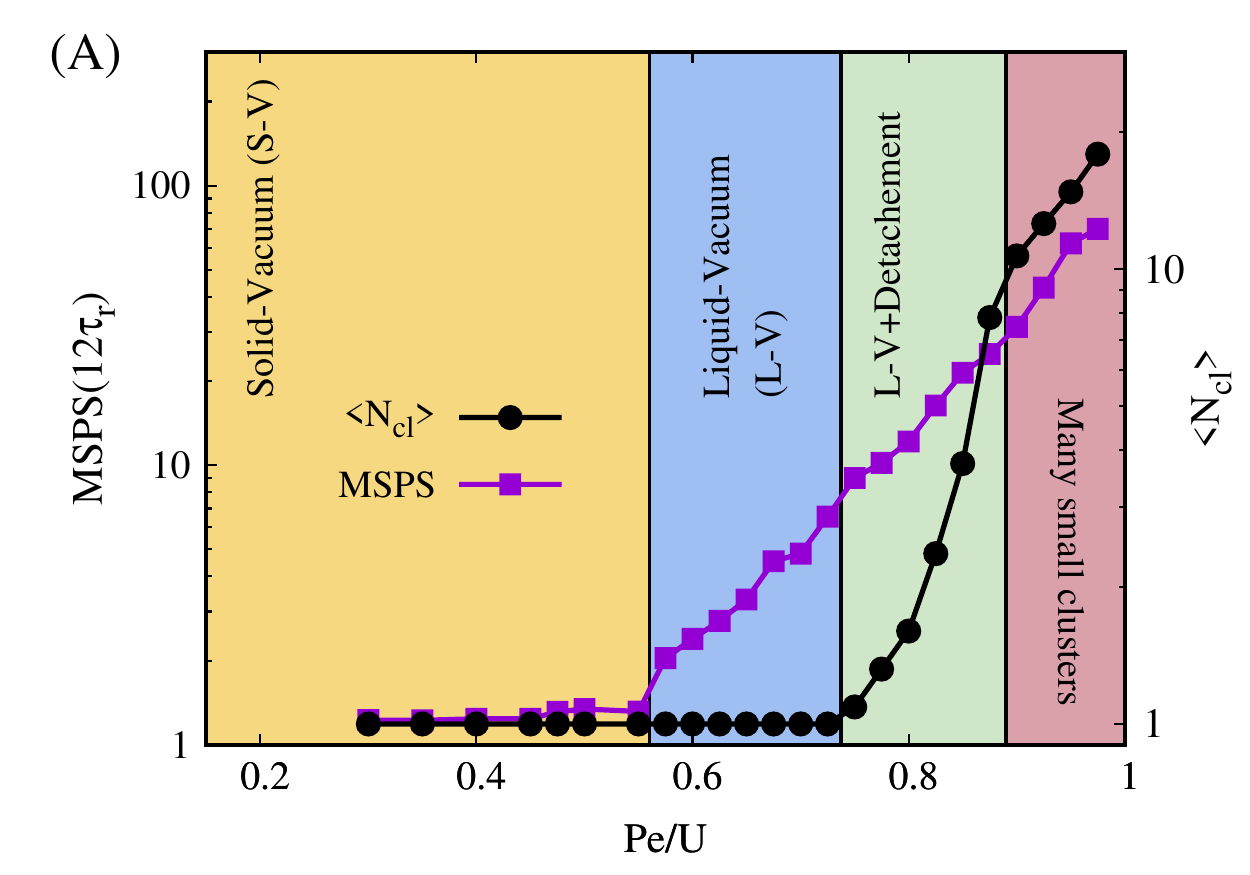}
\includegraphics[width=0.48\linewidth]{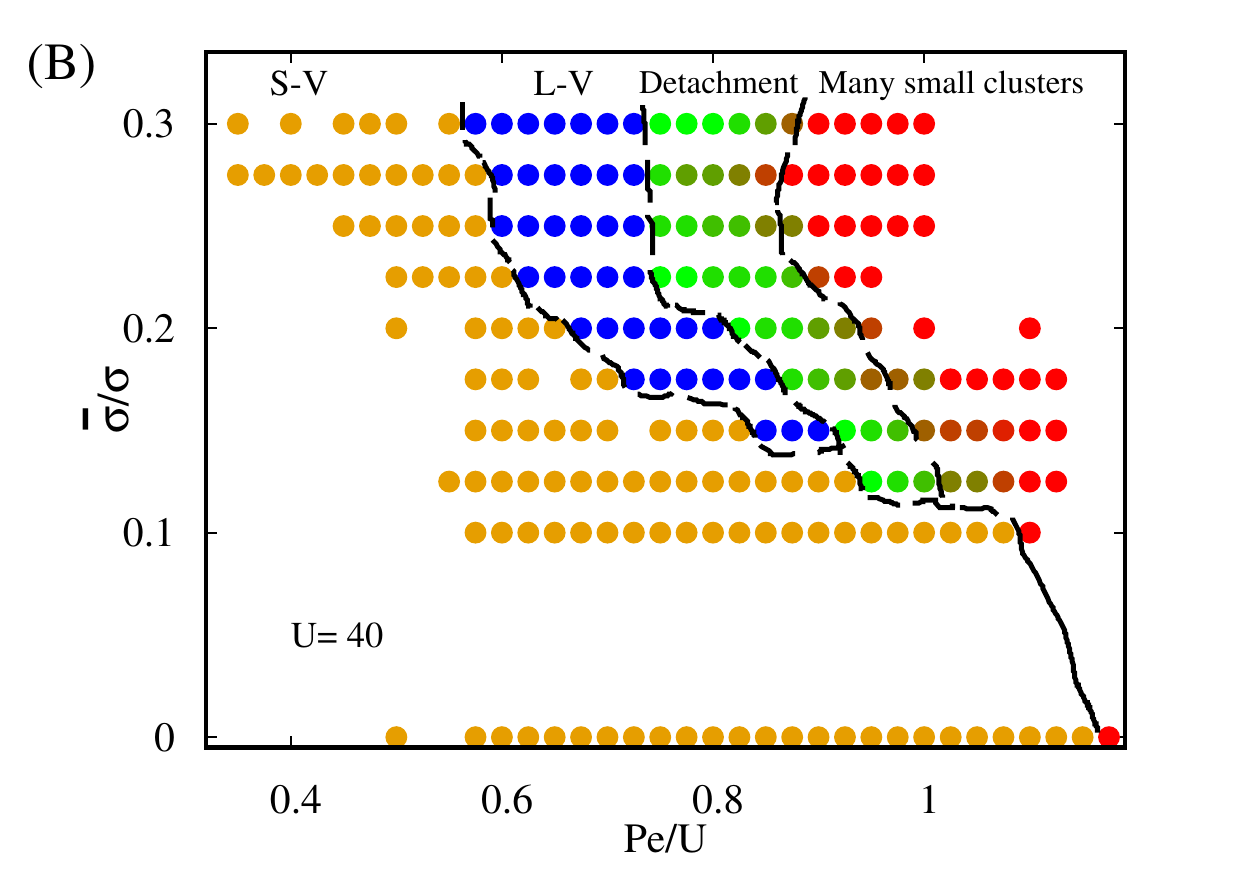}
\includegraphics[width=0.48\linewidth]{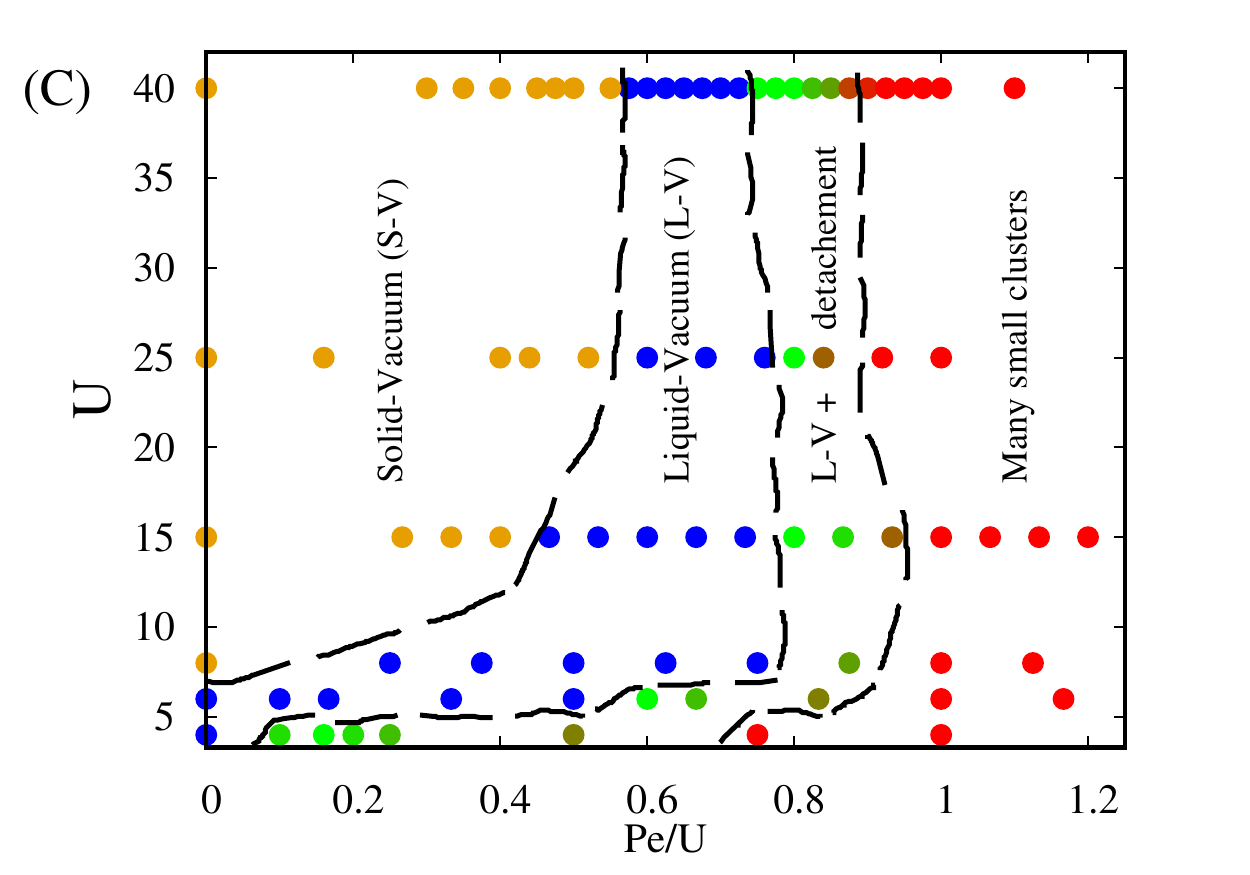}
\includegraphics[width=0.48\linewidth]{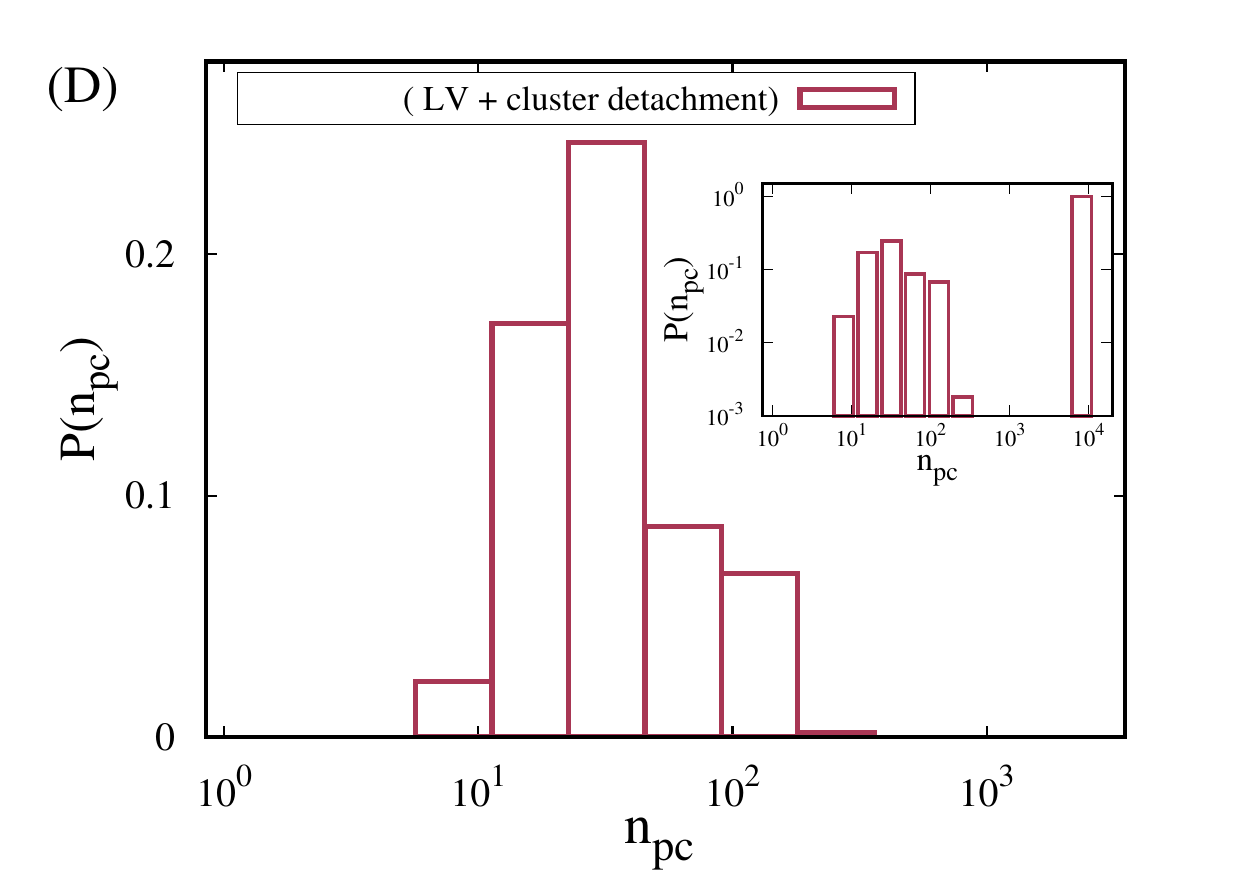}
\caption{(A) "MSPS" (left axis) and $N_{cl}$ (right axis) as a function of $P_e/U$, for fixed $U=40$ with increasing activity.   
changing $Pe$ value. $MSPS(t)$ is calculated at time separation $t=12\tau_r$. The orange area indicates $S-V$ coexistence, the blue 
area $L-V$ coexistence, the green area $L-G$ coexistence with small detached cell clusters in the gas phase, and the red area
a homogeneous phase of many small clusters. Results are for $\bar{\sigma}/\sigma = 0.3$.
(B) Phase diagram as a function of the potential width $\bar{\sigma}$ for fixed $U=40$.
(C) Phase diagram various $U$ as a function of rescaled {P{\'e}}clet number, $Pe/U$, for fixed $\bar{\sigma}/\sigma=0.3$.
(D) Cluster-size distribution of detached clusters, of size $n_{pc}$, at $\bar{\sigma}/\sigma=0.3$, 
$U=40$, and $Pe=32$. Inset: Same data in log-log representations, which also includes the parent colony along with 
the  detached clusters.} 
\label{fig:clstr} 
\end{figure*}

For $Pe \lesssim U$, cells are unable to detach from the colony, and the colony coexists with a cell vacuum outside. 
If $P_e\ll U$ and $U \gtrsim 8$, the system is clearly kinetically arrested, but 
as activity increases, the ``wiggle room'' of the potential allows particles to break the neighbor cage and move, resulting 
in liquid-vacuum (L-V) coexistence. This state of a single cohesive colony is not induced by the initial
conditions of a single circular patch, but also emerges from an initial random distribution of  particles 
due to particle aggregation and cluster coarsening. 
To quantitatively characterize and clearly distinguish  mobile cohesive colonies 
from the kinetically-arrested colonies, we employ the "mean squared particle separation" (MSPS). We choose 
random pairs of cells $m$ and $n$ inside the colony at time $t_p$, which are initially at contact with a 
center-center distance $1.1\,\sigma$, and measure the squared separation of this pair over time. n average over $N_p$ 
such pairs at different initial times $t_p$ yields  
\be\label{eq:4}
MSPS(t)= \frac{1}{N_p}\sum_{N_p} ({\bf r}_m(t_p+t)-{\bf r}_n(t_p+t))^2.
\ee 
A characteristic feature of the arrested dynamics in a solid phase is that particles do not exchange neighbors, so that the MSPS 
plateaus at $MSPS<(1.2 \sigma)^2$. In a fluid phase, particles exchange neighbors at a constant rate, 
and MSPS increases linearly with time (see also Appendix B, Fig.~8). Thus the MSPS is good indicator of fluid-like behavior. 
Here, we choose $MSPS>(1.2\sigma)^2$ at time $t=12\tau_r$ as a definition of fluid-like behavior.
To quantify cohesiveness, we turn to a cluster analysis, where particles are identified to be in the same cluster if their 
distance is less than the cutoff distance $r_{cut}$. 
The condensed phase-vacuum coexistence is then signaled by cluster number $N_{cl}=1$. Figure~\ref{fig:clstr}(A) displays $MSPS(t=12\tau_r)$ 
and $N_{cl}$ as a function of $Pe/U$. For $Pe\lesssim 0.55 U$, the system remains cohesive and solid. As activity increases, 
MSPS increases as well, but the colony remains cohesive, clearly identifying the liquid-vacuum (L-V) coexistence region. 
Further increasing activity ($Pe \gtrsim 0.75 U$) leads to the occasional detachment of small clusters (above a threshold size) 
from the parent colony (see discussion below).  Interestingly, 
occasional cluster detachment is not sufficient to disintegrate the parent colony, as detached cluster can rejoin the parent colony,
which thereby coexists with a gas of small clusters.

Figure~\ref{fig:clstr}(B,C) display different cuts through the phase space, to elucidate the region of stability 
of different regimes.  The results in Fig.~\ref{fig:clstr}(B) show that a minimum width $\bar{\sigma}/\sigma \simeq 0.1$ of 
the potential well is necessary to observe a liquid-vacuum coexisting phase. Thus, the width $\bar{\sigma}/\sigma$ plays a 
crucial role to achieve a cell colony with fluid-like dynamics at strong adhesion. The importance of $Pe/U$ as the relevant 
variable to distinguish two-phase coexistence from a one-phase gas-like region,
is emphasized by Fig.~\ref{fig:clstr}(C), which demonstrates that the boundaries between the different regimes 
occur at $Pe/U\simeq 0.55$, 0.75, and 0.875, for $U\gtrsim 20$. Note that all these boundaries appear at $Pe/U$ values,
which are much smaller than the unbinding threshold $Pe/U\simeq 2.4$ of particle pairs.

For P{\'e}clet numbers $Pe \gtrsim0.75 U$, small clusters are able to detach from
the parent colony. This process can be characterized by the cluster-size distribution $P(n_{pc})$, see Fig.~\ref{fig:clstr}(D). 
The peak of the distribution for clusters in the size range from 10 to 100 indicates that particles escape collectively. 
We do not observe the escape of any single cell from the colony in this regime.
This can be understood from a simple argument, which considers a small semi-circular patch of $n_{pc}$ particles at the
boundary of the colony (see Appendix B, Fig.~9). The patch has an interface with the colony of length proportional to $\sqrt{n_{pc}}$. If 
all particle orientations point in roughly the same direction (outwards), then the patch can unbind when 
$n_{pc} > n_{pc}^* \simeq 12.7 (U/Pe)^2$, i.e. for sufficiently large patch size, a size which decreases rapidly with 
increasing $Pe$ (see Appendix B for details, in particular Fig.~10). 
The probability for all particles in such a cluster 
to be roughly aligned depends on the P{\'e}clet number, as particles move toward the boundary with preferred outward
orientation \cite{Elgeti_EPL2013}. However, the particle mobility in the fluid phase is very small due to the dense 
packing of neighbors, so that the characteristic ballistic motion of ABPs for times less than $\tau_r$ is completely 
suppressed (see Appendix B, Fig.~7).
Therefore, polar ordering is mainly seen at the edge of the colony, see Fig.~\ref{fig:strs}(A). Cluster formation 
therefore arises mainly from the increased mobility of pre-aligned particles at the boundary.

\subsection{Stress Profile - Tensile Colonies}

\begin{figure*}[!htb] 
	\centering
	\includegraphics[width=0.32\linewidth]{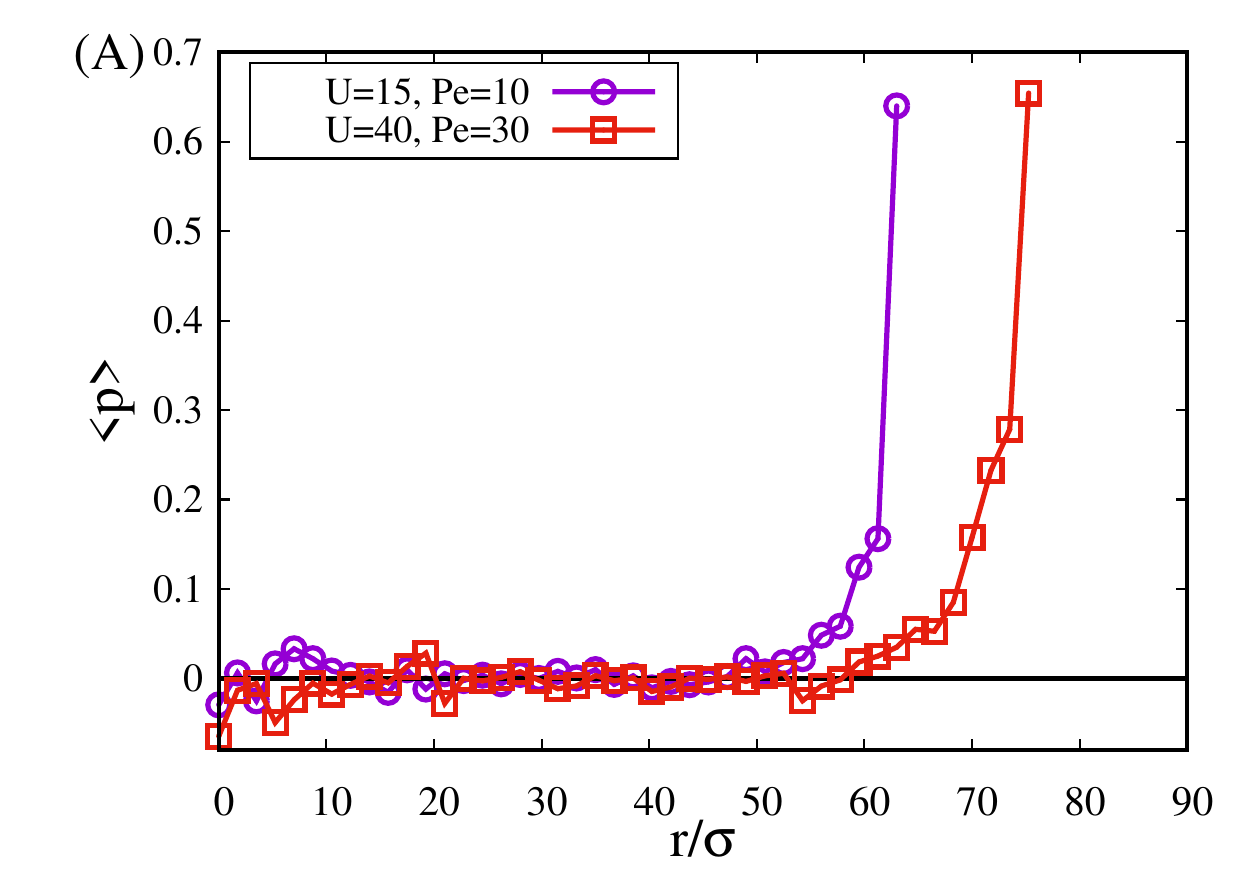}
	 %
	\includegraphics[width=0.32\linewidth]{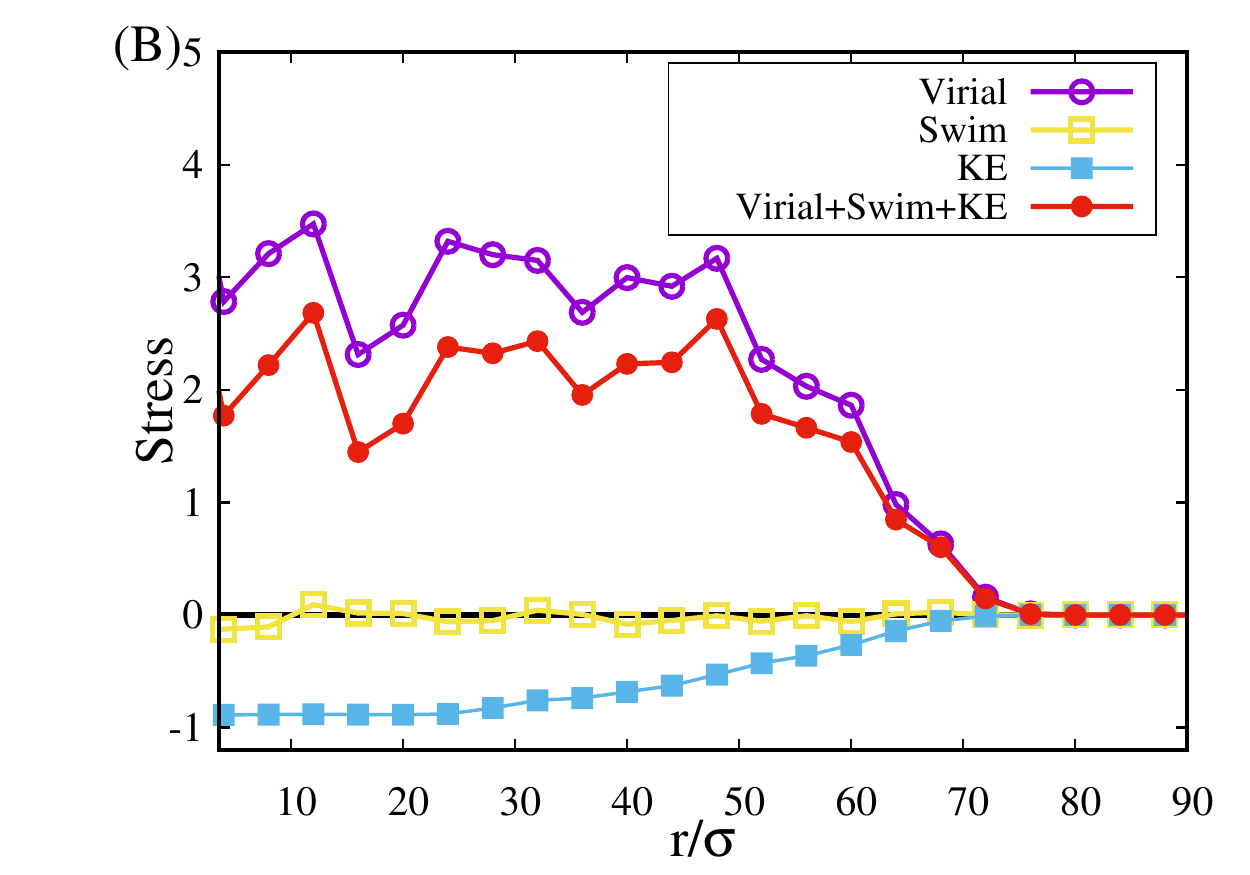}
	   %
	\includegraphics[width=0.32\linewidth]{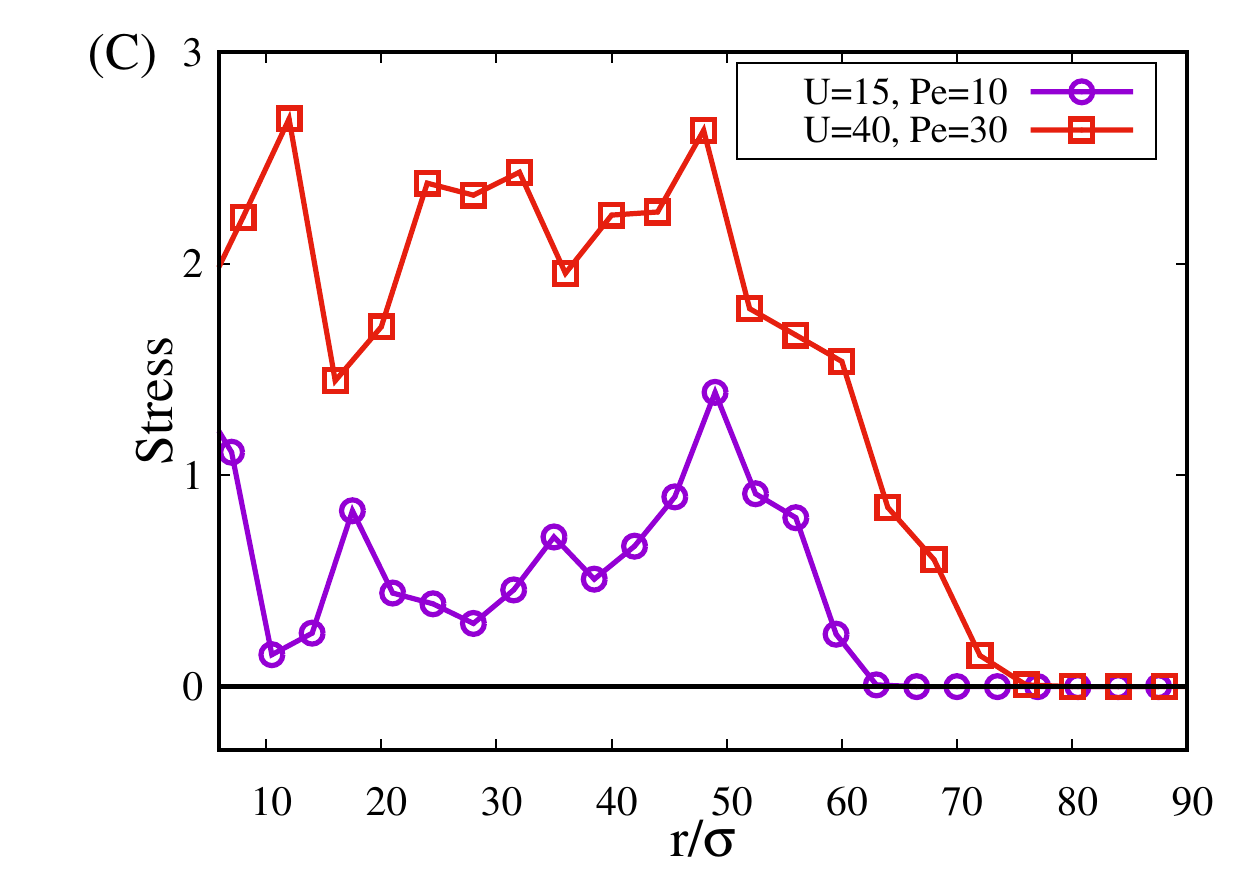}
\caption{(A) Averaged polarization vector of the circular patch at $\bar{\sigma}=0.3 \sigma$ at $L-V$ coexisting state for two 
different activity and 
(B) Different components of the stress at $\bar{\sigma}=0.3\sigma$ and $U=40$, $Pe=30$. 
"Violet" represents the "virial" stress profile, yellow color represents the stress profile due to activity, "sky-blue" 
represents the kinetic contribution of the stress profile and "red" represents the sum of all these three contributions. 
(C) The total stress profile for different adhesive strength in $L-V$ coexisting states at $U=15, 40$. The stress starts 
generated at the boundary region of the colony and outside of it, the stress vanishes.}
\label{fig:strs} 
\end{figure*}

\begin{figure}[ht]
\centering
\includegraphics[width=0.95\linewidth]{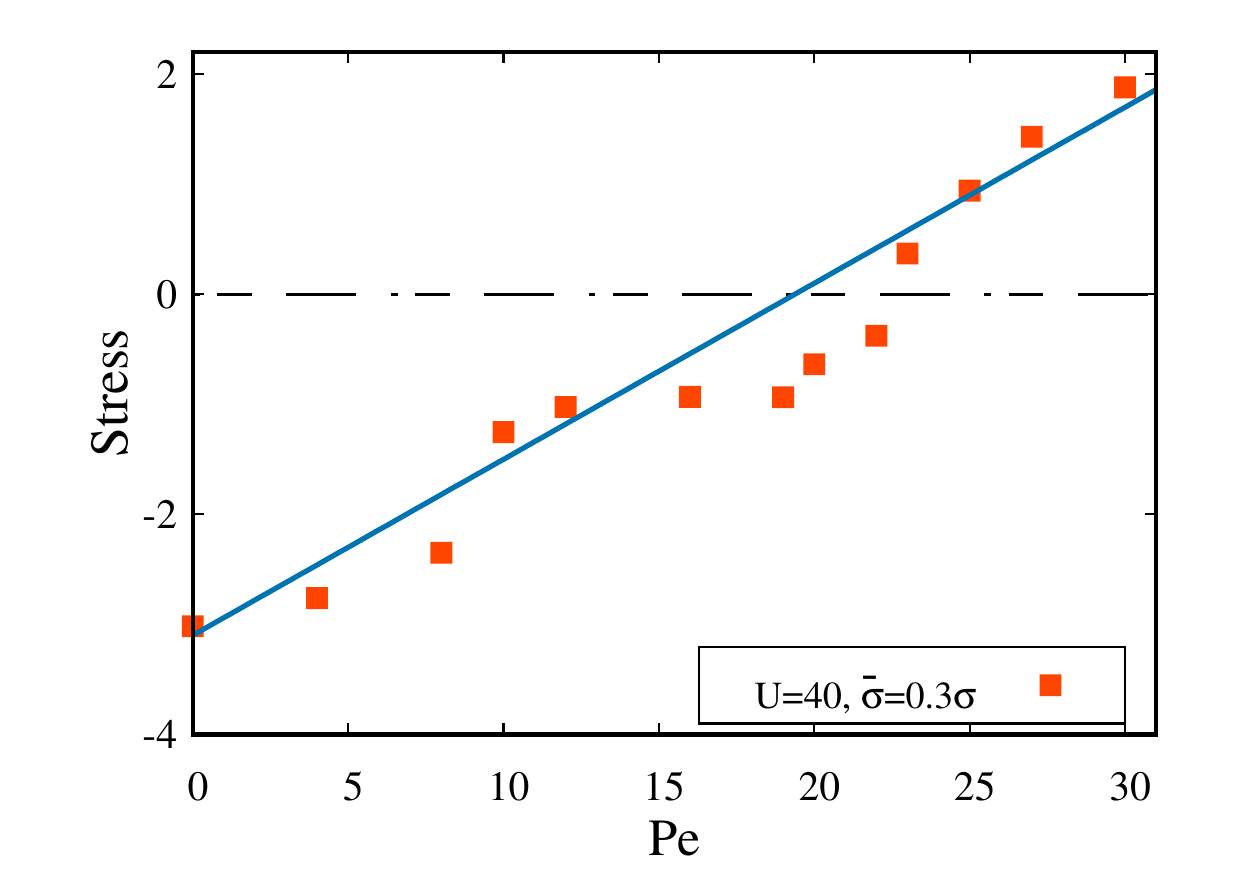} 
\caption{Total central stress calculated in the area starting from the center to a radius $30\sigma$ as a 
function of activity, $Pe$, for $U=40$ and $\bar{\sigma}=0.3\sigma$. The transition from S-V to L-V
coexistence occurs at $Pe=23$ (compare Fig.~\ref{fig:clstr}(C)), which essentially coincides with 
P{\'e}clet number where central stress changes sign.}
\label{fig:strs-u40}
\end{figure}

To gain a better understanding of the properties of the liquid-like cohesive colony, we analyze the polarization field 
of the active force and the stress profile. Figure~\ref{fig:strs}(A) shows that the averaged local polarization 
is zero inside the colony, but points outwards at the boundary. This is in contrast to what is typically found for 
motility-induced clustering \cite{marchetti-PRL,fily-2014,wysocki_EPL2014,ignacio-PRL-2018,Das_SREP2019}.
The reason is that the attractive interactions keep outward-oriented particles at the boundary of the 
colony -- which would otherwise move away -- combined with the fluidity of the colony which allows local particle
sorting near the boundary, similar to the behavior of isolated self-propelled particles in confinement 
\cite{Elgeti_EPL2013,Fily_SM_2014}.

The alignment of motility forces at the boundary should lead to an increase of tensile stress. For the liquid colonies in 
coexistence with the vacuum phase, we find significant tensile stress in the center (see Fig.~\ref{fig:strs}(C)). 
As expected from force balance, the stress is nearly constant inside the colony, but rapidly decreases in the boundary 
region where the tension is generated. The total stress in the colony has three contributions: the inter-particle force, 
kinetic contribution, and swim stress. At liquid-vacuum coexistence, the inter-particle contributions plays the dominant role, 
whereas the swim stress is comparatively small (Fig.~\ref{fig:strs}(B)). Figure~\ref{fig:strs-u40} shows the dependence of 
the total central stress in the colony on the activity $Pe$. In the solid regime, the central stress is negative due to 
the passive surface tension, resulting in a Laplace pressure proportional to $U$. Increasing activity leads to more 
liquid-like consistency, facilitating enhanced outward particle orientation at the edge, and hence tensile stress. 
Interestingly, the stress is found to be a linearly increasing function of $Pe$ over the whole investigated 
range, $0< Pe \le 30$, i.e. both in the solid and liquid regime of the colony. This indicates that the enhanced particle 
sorting occurs mainly near the edge, and an increased edge mobility exists already in the solid phase near the S-L phase boundary.
A tensile stress in cell colonies is observed similarly in experiments, 
where the average stress within a spreading cell sheet increases as a function of distance from the 
leading edge \cite{trepat-nat}. In a quasi-one-dimensional (rectangular channel) geometry, the total stress in the 
colony is obtained by integration of the net active forces, and increases from zero outside to a finite tensile stress 
in the center (see Appendix B, Fig.~11). 


\subsection{Coordinated Motion -- Motion Alignment}
\label{sec:coor}

\begin{figure}[ht]
\centering
\includegraphics[width=0.93\linewidth]{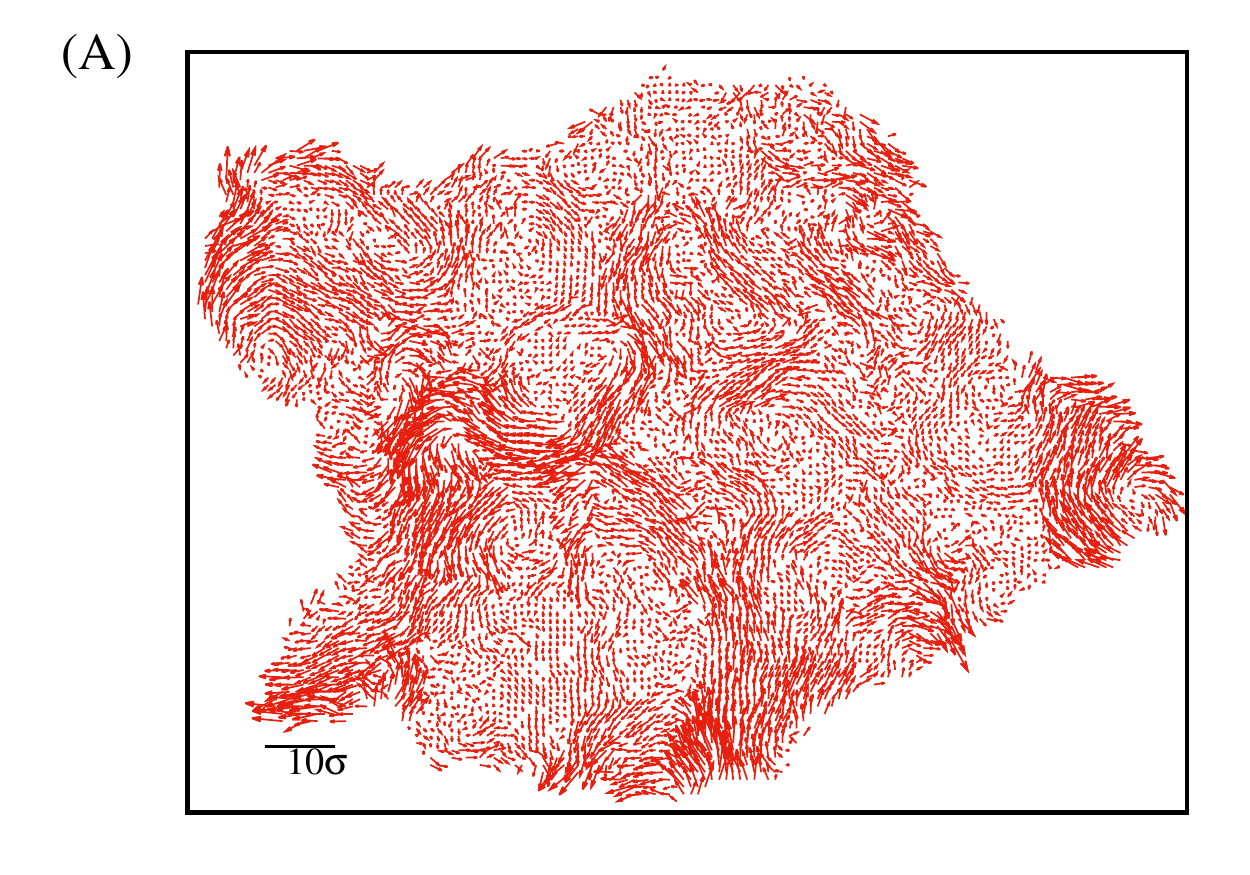} 
\includegraphics[width=0.93\linewidth]{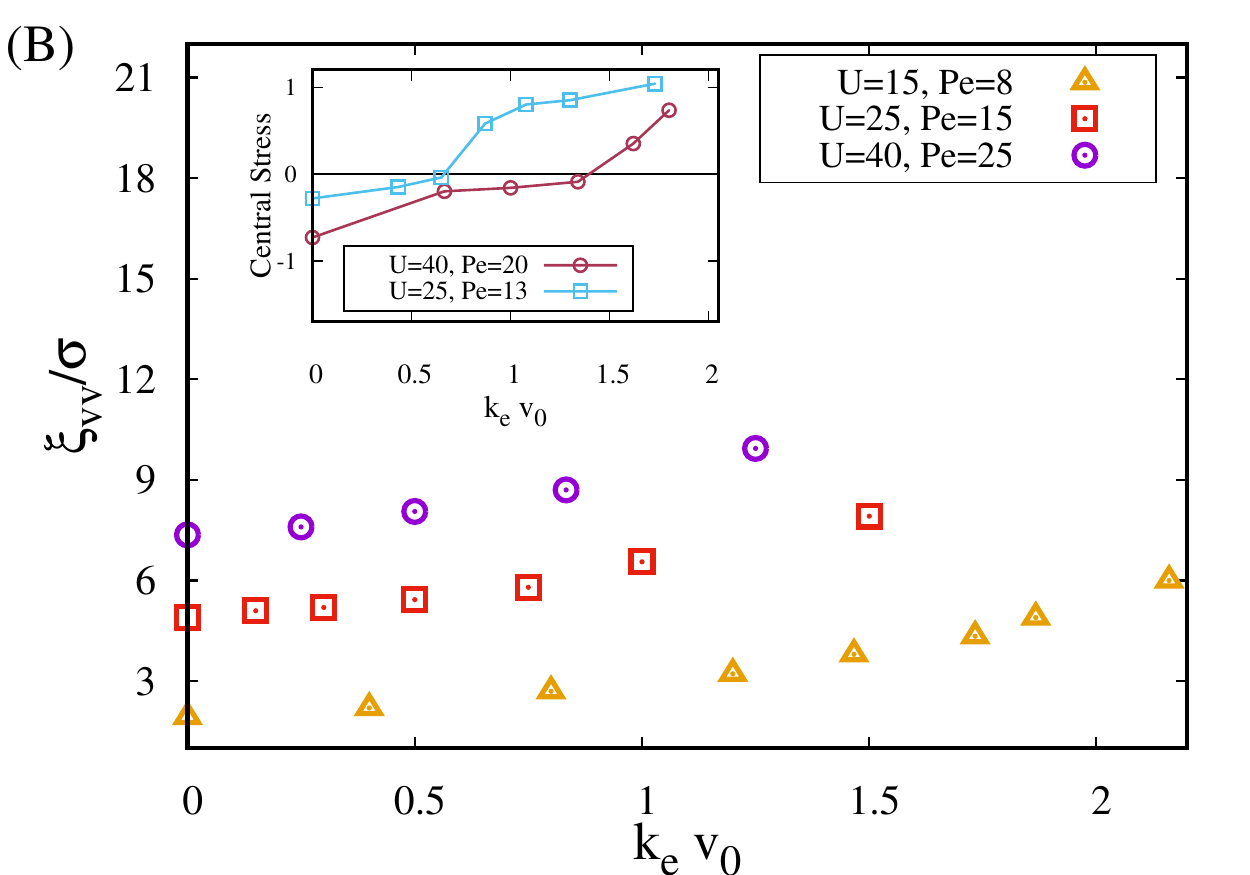} 
\caption{Dynamics of cell colony with orientation-velocity alignment interactions.
(A) Velocity fluctuation field of the full cell colony for 
$U=40$, $Pe=25$ and $k_e v_0=1.25$ at $\bar{\sigma}=0.3\sigma$, which displays prominent swirls in the bulk and finger-like 
structures at the edge. See also Appendix B, Fig.~13.
(B) Characteristic correlation length $\xi_{vv}$, extracted from velocity correlation, as a function of alignment 
strength $k_e$, for different attractive interactions $U$ and activities $Pe$, as indicated, with $\bar{\sigma}=0.3$. 
The colony is in the fluid state for $k_ev_0=0$.
Inset: Variation of total central stress as a function of alignment strength $k_e$, for adhesive strength 
$U=25, 40$ ($\bar{\sigma}=0.3$). 
For both data sets, the colony is in the solid state for $k_e=0$, and transits 
to a liquid state at $k_ev_0 \simeq 0.65$ and $k_ev_0 \simeq 1.3$ for $U=25$ and $U=40$, respectively.} 
\label{fig:prf}
\end{figure}

In experimental observations, long-range velocity correlations are often visible in the bulk of spreading epithelial 
sheets \cite{angelini-10, angelini-11}. ABPs with adhesion display significant velocity correlations even without 
explicit alignment interactions (see Appendix Fig.~13, and Refs.~\cite{garcia-2015,hakim-plos}). However, as ABPs 
display independent orientational diffusion, it is evident that realistic long-range correlations require some type 
of velocity alignment. 
We employ a local velocity-orientation alignment mechanism, in which cell propulsion direction (=cell polarity) 
relaxes toward the instantaneous cell velocity, resulting from the forces induced by its neighbors
\cite{szabo-06,basan-2013,dauchot-2015}, as introduced in Eq.~\ref{eq:2}.  
Without alignment, correlations arise from a small group of cells pointing in the same direction by chance, and thus 
moving collectively more easily and furthermore dragging other cells along. The alignment interaction stabilizes and 
enhances this effect. In presence of velocity alignment, the velocity field shows an enhanced coordinated motion with 
prominent swirls in the bulk of the colony and fingering at the edge (see Fig.~{\ref{fig:prf}}(A)).

We quantify the spatial correlations by the velocity-velocity correlation function
\be
C_{vv}(r)=\Bigg \langle \frac{\sum_{r_i} \delta{\textbf{v}}({\textbf{r}}_i) 
                            \cdot \delta{\textbf{v}}({\textbf{r}}_i + {\textbf{r}})}
{\sum_{r_i}\delta{\textbf{v}}({\textbf{r}}_i) \cdot \delta{\textbf{v}}({\textbf{r}}_i)} \Bigg \rangle,
\ee
as a function of distance $r$, where the brackets denote an average over all directions and time. Here, velocities are measured 
relative to the average velocity $\bar{\textbf{v}}$, of the whole colony, 
i.e. $\delta{\textbf{v}(\textbf{r})}={\textbf{v}(\textbf{r})}-\bar{\textbf{v}}$, 
to avoid finite-size effects. The correlations decay exponentially with a characteristic length scale, $\xi_{vv}$ 
(see Appendix B, Fig.~14).
Figure~\ref{fig:prf}(B) displays the correlation length $\xi_{vv}$ as a function of alignment strength $k_e$ 
for various adhesive interactions.  
For fixed adhesion and activity, increasing alignment strength $k_e$ facilitates a transition from the solid to the liquid state 
of the colony. Furthermore, the alignment coupling leads to stronger correlations, as indicated by the monotonic increase of
$\xi_{vv}$ with $k_ev_0$, and thus to swirls and fingers. 
Eventually, fingering is so strong that clusters detach, and the colony is no longer cohesive. However, 
correlation lengths up to $\xi_{vv}=10\sigma$ can be achieved, quite comparable to the 5 to 10 times cell size obtainable in 
experiments \cite{angelini-10,garcia-2015}.  Also, the tensile stress at the colony center increases (see inset of Fig.~\ref{fig:prf}(B))  
and becomes positive at sufficiently large $k_ev_0$. The critical alignment strength $k_ev_0$, where the colony is liquefied and the 
tensile stress becomes positive, increases with attraction strength $U$.


\section{Conclusions and Outlook}
\label{sec:end}

We have presented a minimal model for the fluidization of cell colonies, which consists of active Brownian particles 
with adhesion. An attractive potential with increased basin width yields non-equilibrium structures, phase behavior and dynamics, 
which capture relevant features of biological cell colonies. 
The main observation is that for moderate adhesion and propulsion, the system exhibits liquid-vacuum coexistence, i.e. all 
particles agglomerate into a single colony displaying liquid-like properties, while the outside remains devoid of any particles. 
This is reminiscent of {\em in vitro} experiments of MDCK colonies, where cells show strong motion, while remaining perfectly 
cohesive. Furthermore, the fluidity of the colony in our model results in outward ordering of particle orientations at the edge, 
thus leading to tension in the colony. This is consistent with the results of traction force microscopy, which show 
that MDCK colonies are typically under tension \cite{trepat-nat}. Our model demonstrates that no alignment interaction or 
growth mechanism need to be evoked to explain such tensile forces -- the motility of the cells combined with liquid properties 
of the colony suffice. 
As motility force increases, particles start to detach from the parent colony, however not as  single cells but collectively 
in small clusters of cells. 
Finally, we have demonstrated how velocity-polarity alignment can further enhance fluidity, tension, and fingering of the 
colony, and collective cell detachment. 
Indeed, with velocity-polarity alignment, simulations look very reminiscent of real MDCK colonies, displaying strong 
fingering at the edge, high tension and long-ranged velocity correlations.

Our model also provides a tentative explanation for another biological phenomenon. 
When metastatic cells detach from a tumor, they typically detach collectively, as small groups of five cells or large
aggregates \cite{gilmour-nature,rajasekaran-2006,vignjevic-2015}, into the stroma and migrate to reach blood or lymph 
vessels.  At the edge of the liquid-vapor region of our model, particles show exactly this type of behavior; the 
colony is no longer perfectly cohesive, but clusters of cells begin to detach.

An interesting next question is how these results will be affected by cell growth.
Of course, if growth is slow, the dynamics will be independent of growth and the phenomenology will be unchanged.
However, when time scales of growth and motion become comparable, novel phenomena may arise.

\begin{acknowledgments}{Financial support by the Deutsche Forschungsgemeinschaft (DFG) through the priority program SPP1726 
``Microswimmers -- from single particle motion to collective behavior'' is gratefully acknowledged. 
The authors also gratefully acknowledge the computing time granted
through JARA-HPC on the supercomputer JURECA \cite{JURECA_2018} at Forschungszentrum J\"ulich.}

\end{acknowledgments}



\appendix

\section{Attractive Brownian Particles: Model and Analysis of Simulation Data}
\subsection{Simulation parameters}
\begin{figure}
	\centering
	\includegraphics[width=0.37\textwidth]{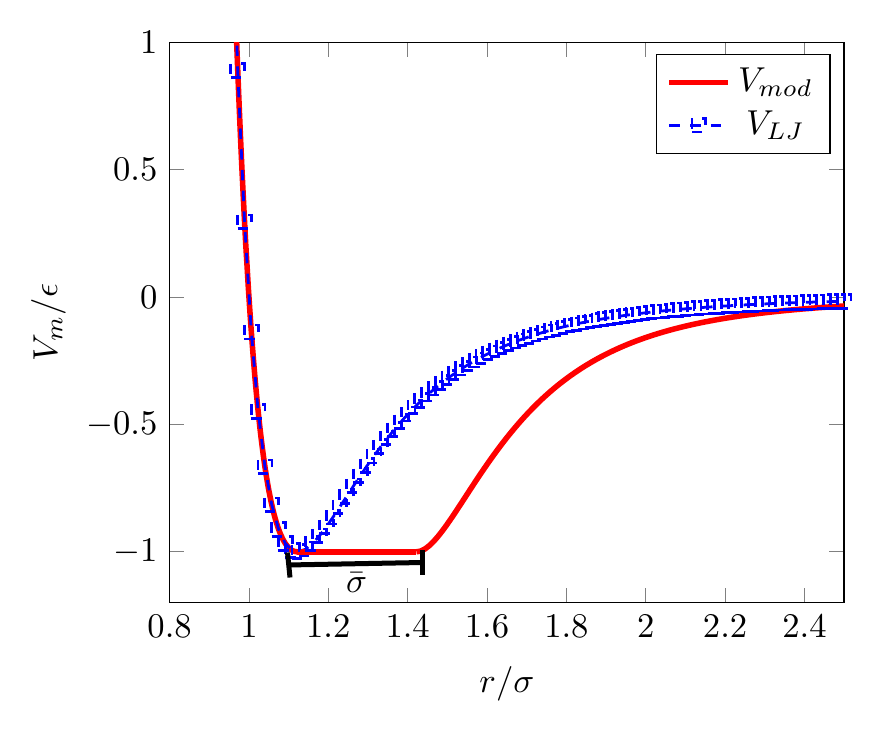}
	\caption{Modified Lennard-Jones (LJ) potential (red colored curve) is modified by inserting a plateau of 
		width $\bar{\sigma}$ at the minimum of this potential. The usual LJ potential is shown by blue crosses.}
	\label{fig:pot}
\end{figure}

For numerical implementation of our model, we use the LAMMPS molecular simulation package, with in-house 
modifications to describe the angle potential and the propulsion forces, as described in the main text. 
The system consists of $N = 7851$ particles (cells) in a 2D square simulation box of size 
$L_x = L_y = 150\sigma$ with periodic boundary conditions, unless noted otherwise. With velocity-orientation-alignment 
interaction, the simulation is carried out in a box of size $L_x=L_y=250\sigma$. For the extended LJ interaction potential, 
see Fig.~\ref{fig:pot}, we use the cut-off distance $r_{cut}=2.5\sigma$. For numerical efficiency, 
we chose a finite mass $m=1$ and drag coefficient 
$\gamma=100$ such that the velocity relaxation time $m/\gamma$ is much smaller than all physical time scales. 
The equations of motion are integrated with a velocity Verlet algorithm, with time step $\Delta t=0.001$. 
Each simulation is run for at least $11\times 10^7$ time steps, with rotational diffusion coefficient $D_r=0.03$ this 
corresponds to a total simulation time longer than $3000 \tau_r$, where $\tau_r$ is the rotational decorrelation time.

\subsection{Polarization Vector}
We define the spatial-temporal average polarization $p$ for the quasi-circular cell colony by the projection of the 
orientation vector $\bf \hat n$ of the particle on the radial direction from the center of mass of the colony, i.e.,  
\be
\langle p (r',t) \rangle = \sum_{i=1}^{N} ({\bf \hat{n}}_i \cdot {\bf \hat{r}}_i^{\prime}) \, 
\delta(r' - |{\bf r}'_i|) / \sum_{i=1}^{N} \delta(r' - |{\bf r}'_i|) 
\ee
where ${\textbf{r}}_i^{\prime}= {\textbf{r}}_i -{\textbf{r}}_{cm}$, and ${\textbf{r}}_{cm}$ is the center-of-mass 
position at a particular time $t$. Here, $\delta(r)$ is a smeared-out $\delta$-function of width $\sigma$. 
$\langle p \rangle$ is further averaged over time.

\subsection{Stress calculation}

In ABP systems with short-range repulsion, it has been shown that the pressure is a 
state function, depending only on activity, particle density, and interaction potential, but {\em not} on the interaction 
with confining walls \cite{Takatori-PRL-2014,Solon-NatPhys-2015,rl-jack-pre-2016,Das_SREP2019}. 
In comparison to passive systems, activity implies a new contribution to pressure, which is called the swim pressure.
The calculation of the local stress in an ABP system is a matter of an ongoing debate, which mainly concerns 
the form of the active term. We follow Ref.~\cite{Das_SREP2019}, and define the stress in a volume $\Delta V$ by 
\be\label{eq:6}
\begin{aligned}
	\Delta V \Sigma_{\alpha \alpha} =
	& \- \sum_{i=1}^{N}m\langle\dot{\bf r}_i^2\Lambda_i\rangle 
	-\frac{\gamma}{\gamma_R}\sum_{i=1}^{N}\langle v_0 {\bf n}_i \cdot \dot{\bf r}_i\Lambda_i\rangle \\
	& \-\frac{1}{2}\sum_{i=1}^{N}\sum_{j=1}^{N}\langle \lambda_{ij} {\bf r}_{ij} \cdot {\bf F}_{ij} \rangle
\end{aligned}
\ee
where $\dot{\bf r}_i$, ${\bf F}_i(i=1,...,N)$ denote the velocity and force of particle $i$, respectively.  
${\bf F}_{ij}$ represents the pair wise interaction between particle $i$ and $j$ and 
${\textbf{r}}_{ij}={\textbf{r}}_i-{\textbf{r}}_j$. Here $\Sigma_{\alpha \alpha}$ are the diagonal stress-tensor components. 
$\gamma_R$ is the damping factor which is related to the rotational diffusion coefficient as $\gamma_R=2D_r$. 
$\Lambda_{i}$ determines the volume $\Delta V$, where $\Lambda_i(\textbf{r})$ is unity when particle $i$ 
is within $\Delta V$ and zero otherwise. 
$\lambda_{ij}$ denotes the fraction of the line connecting particle $i$ and $j$ inside of the volume $\Delta V$. 
The first and the last term in ~\eqref{eq:6} are the classic kinetic contribution and the contribution 
of inter-particle interactions. The second term denotes the active-force contribution in the stress calculation. 
Notably, in the fluid-vacuum state we are focusing on in this work, the active stress component is negligible compared 
to the inter particle interaction contribution. 
This is in line with results for the pressure contributions in repulsive ABP systems at coexistence between
a high-density and a low-density phase, where the swim pressure in the high-density phase is negligible \cite{Das_SREP2019}.

\section{Supporting Considerations and Results}

\subsection{Particle Mobility in Vacuum, Gas, and Fluid Phases}

\begin{figure}
	\centering
	\includegraphics[width=0.9\linewidth]{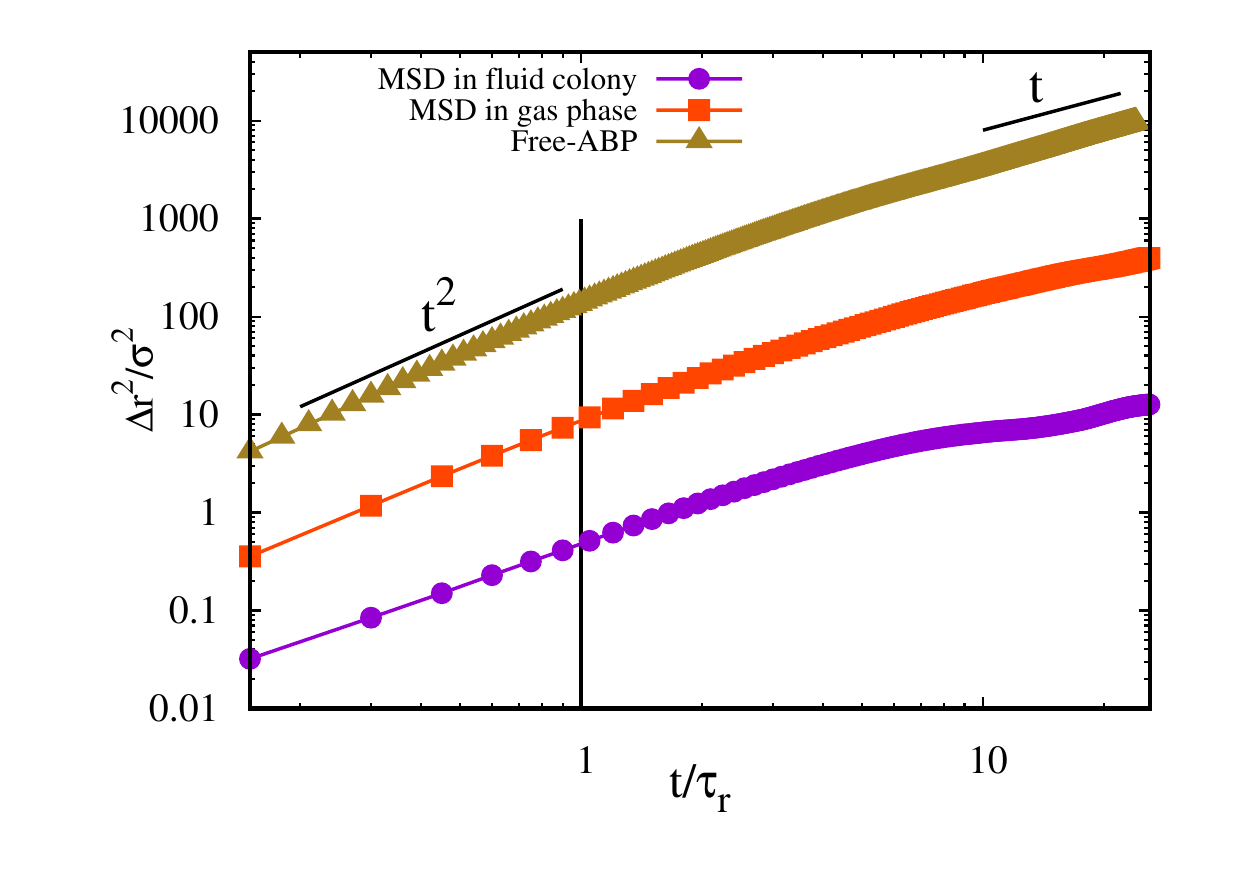}
	\caption{
		Mean square displacement (MSD) in three different states: (i) In the bulk of "fluid-like" colony 
		(purple circles) with $U=40$ and $Pe=30$, 
		(ii) in the "gas" phase (red squares) $U=40$ and $Pe=41$ and packing fraction $\phi=0.274$, and (iii) in the "vacuum" phase 
		(brown triangles) with $Pe=41$. } 
	\label{fig:msd}     
\end{figure}

A single, isolated ABP has a characteristic mean square displacement, with ballistic motion (MSD $\sim v_0^2 t^2$)
at short times $t < \tau_r$, and a diffusive motion (MSD $\sim v_0^2 \tau_r t$) for $t > \tau_r$ \cite{Howse-PRL-2007}.
This is the behavior we observe in the vacuum phase, see Fig.~\ref{fig:msd}. A very similar behavior is found in the
gas phase, at packing fraction $\phi=0.274$, but now the particle velocity is significantly reduced due to frequent
collisions with other particles, while the crossover time $\tau_r$ remains unaffected. However, the behavior
changes drastically in the fluid-like phase, where the collisions and attractive interactions completely 
suppress the ballistic regime, see Fig.~\ref{fig:msd}. 

\begin{figure}
	\centering
	\includegraphics[width=0.4\textwidth]{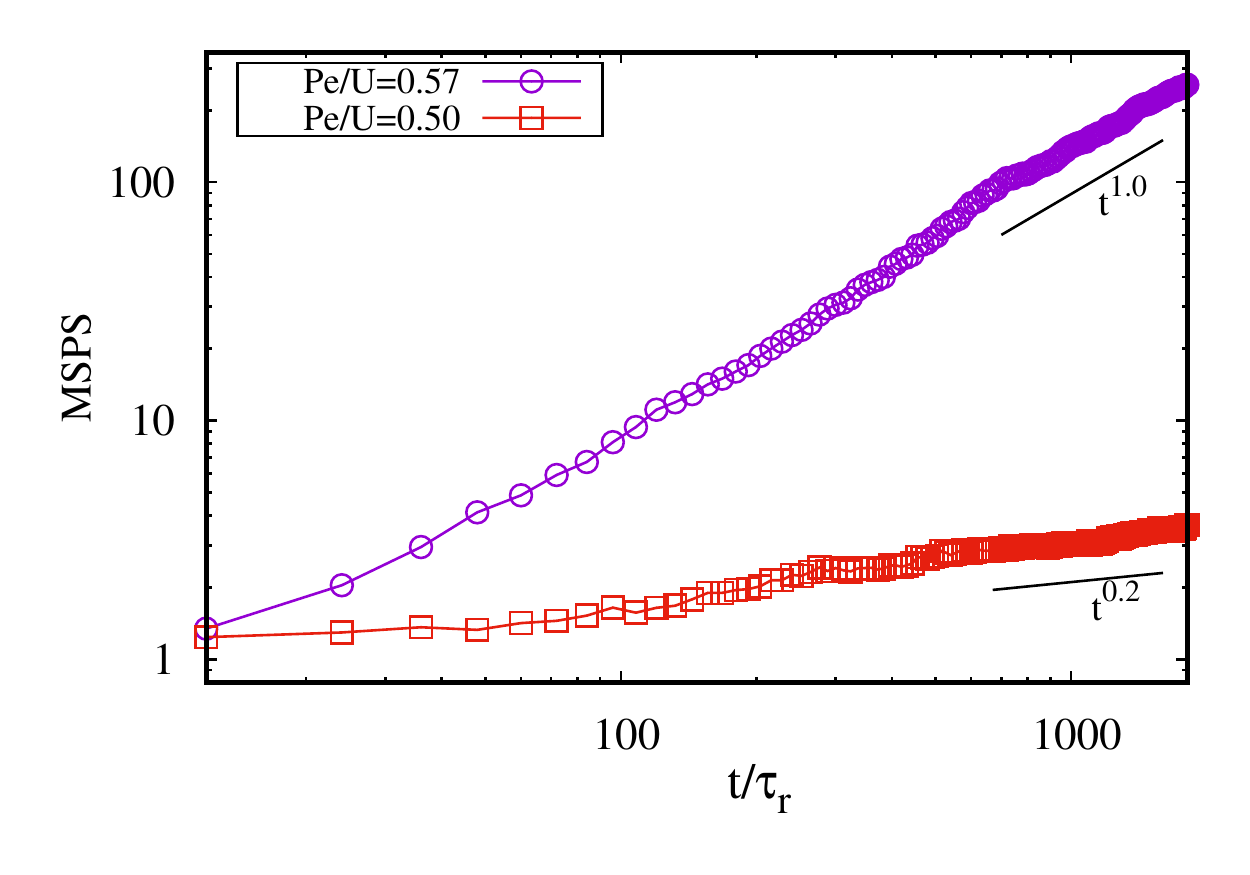}
	\caption{Mean squared particle separation ($MSPS$) as a function of time for two different $Pe$ numbers at $U=40$ and
		$\bar{\sigma}=0.3$. At $Pe/U=0.5$, $MSPS$ remains constant at short times, indicative of no neighbor exchange,
		and shows sub-diffusive behavior at longer time. However, at $Pe/U=0.57$, the time evolution of $MSPS$ shows a long-time 
		linear behavior, indicative of constant neighbor exchange and liquid-like behavior.}
	\label{fig:msps-2}
\end{figure}

A very similar dynamic behavior is observed in the mean squared particle separation (MSPS), 
see Fig.~\ref{fig:msps-2}.  The time dependence in the fluid phase is dominated by linear diffusion behavior, while 
in the solid (jammed) phase it is strongly sublinear. 

\subsection{Minimum Cluster Size for Detachment}

For our model system, particles can only escape from the parent colony in the form of a small cluster at larger adhesive 
strength ($U \geq 10 $). Let us consider the following simplified model. A semi-circular cluster of particles 
has formed at the colony edge, where all particle orientations are aligned and are pointing outwards 
normal to the interface. In this idealized situation, we can address the question about the number of particles
in the cluster, and the ${\rm Pe}$ number required to separate the cluster against the adhesive force from 
the parent colony.

\begin{figure}
	\centering
	\includegraphics[width=0.45\textwidth]{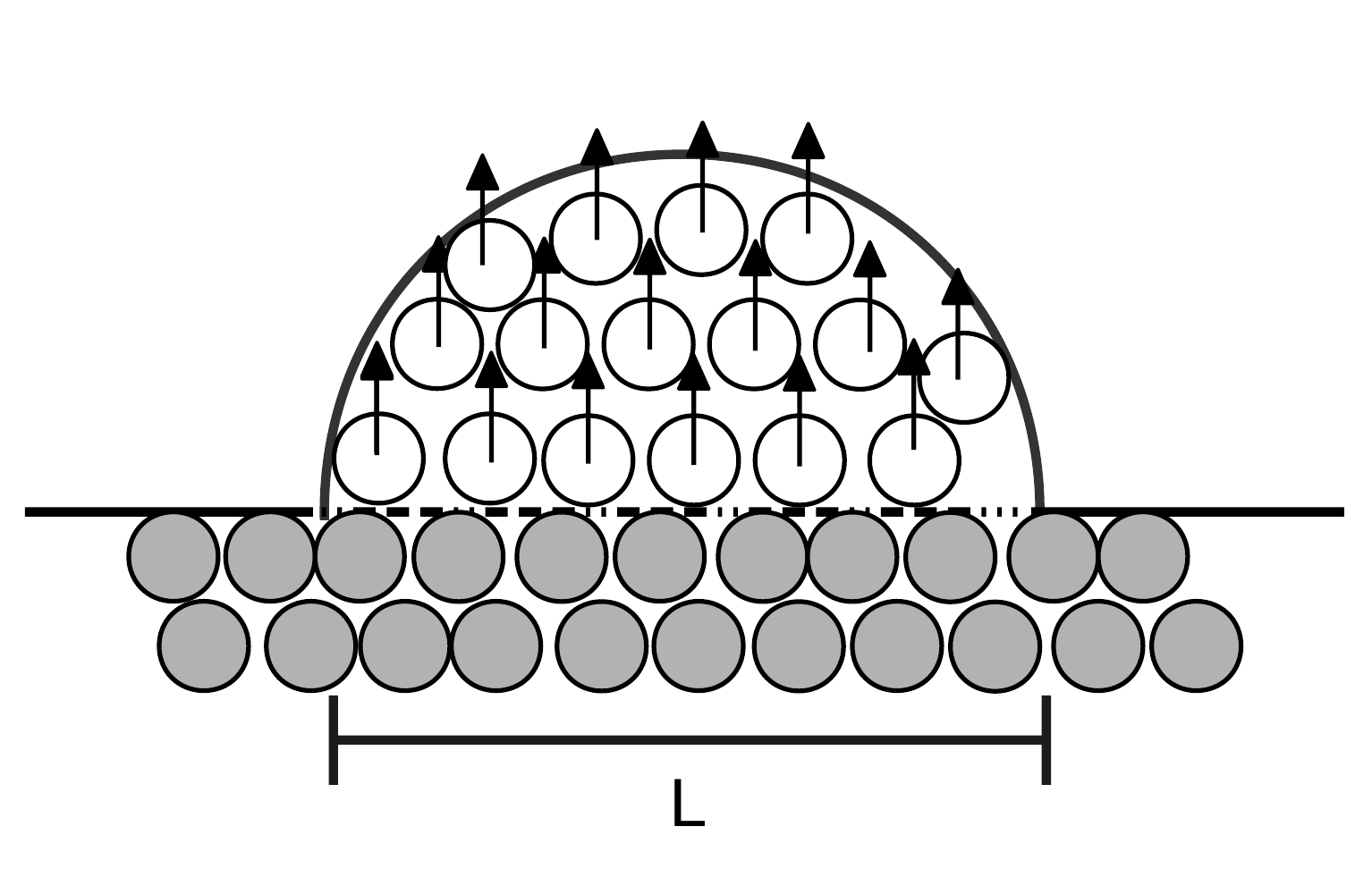}
	\caption{A semi-circular cluster of particles formed at the colony edge. All particles in this region are 
		assumed to be aligned and to be oriented in the outward direction, normal to the interface, as indicated by the arrows. 
		The length of the interface between parent colony and detaching cluster is $L$.}
	\label{fig:toy-mdl}     %
	\end{figure}

\begin{figure}
	\centering
	\includegraphics[width=0.4\textwidth]{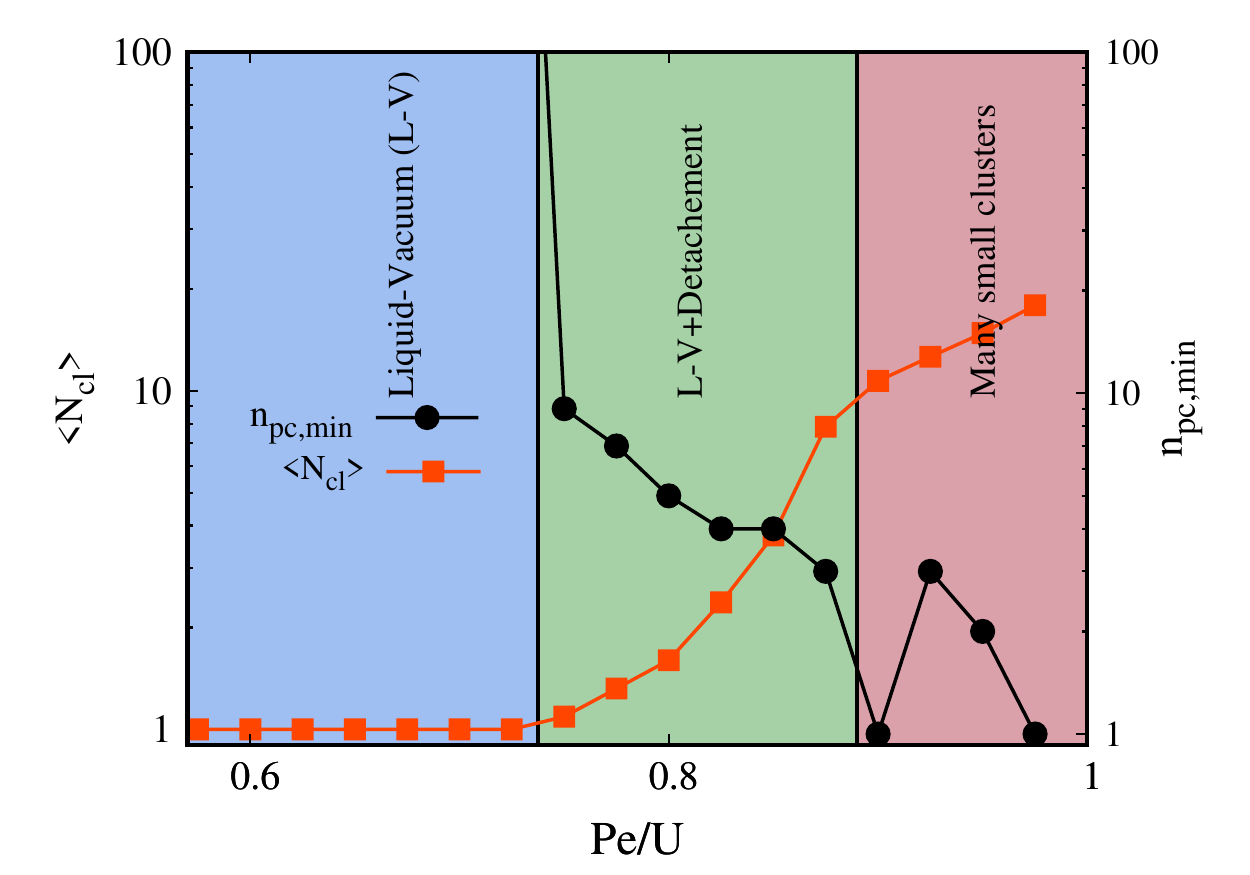}
	\caption{Average cluster size $\langle N_{cl} \rangle$ (left axis) and size of smallest cluster $n_{pc,min}$ 
		(right axis) as a function of $Pe/U$, for fixed $U=40$ with increasing activity ($\bar{\sigma}/\sigma=0.3$). 
		The blue area indicates $L-V$ coexistence, the green area $L-V$ coexistence with small detached clusters 
		in the gas phase, and the red area a homogeneous phase of many small clusters. }
	\label{fig:clstr-sml}
\end{figure}

Let $n_{pc}$ be the number of particles in the cluster. For packing in a roughly triangular lattice with
lattice constant $a=\sigma + \bar{\sigma}/2$, this implies a cluster radius $R_{cl}/a = (\sqrt{3}/\pi)^{1/2} n_{pc}^{1/2}$.
The length of the interface between cluster and parent colony is $L=2R$, see Fig.~\ref{fig:toy-mdl}. 
Along the interface, there are $L/a$ bonds between particles on both sides of the interface, which generate 
the same maximum force as for the detachment of a single particle, where bond breakage occurs at $Pe/U=2.4$ (see main text).
Thus, we have the force balance for the cluster break-off, 
$n_{pc} Pe = 2.4 (L/a) U = 4.8 (\sqrt{3}/\pi)^{1/2} U n_{pc}^{1/2}$,
which implies
\begin{equation}
\label{eq:clustersize_min}
n_{pc,min} = \frac{\sqrt{3}}{\pi} \left( \frac{4.8 U}{Pe} \right)^2.
\end{equation} 
Hence, with increasing activity $Pe$ at constant adhesive strength $U$, the minimal size of detached clusters is expected to
decrease rapidly. Figure~\ref{fig:clstr-sml} shows simulation results for the number of particles present in the smallest cluster 
$n_{pc,min}$ as a function of $Pe/U$. This demonstrates that beyond the liquid-vacuum region, $n_{pc}$ rapidly decreases  
with increasing activity $Pe$ and eventually reaches a "gas-like" phase, where single-particle detachment from the parent colony, 
i.e. $n_{pc,min}=1$, is observed.    

Furthermore, we can use Eq.~(\ref{eq:clustersize_min}) to estimate the cluster size when cluster break-off first
becomes possible, at $Pe/U \simeq 0.75$, which is about $n_{pc}\simeq 20$, in reasonable agreement with the
lower cutoff of the cluster-size distribution in Fig.~2(D) of the main text.

\subsection{Probability Distributions of Aligned Particle Clusters at Colony Edge}

For randomly oriented particles, the probability to find a cluster in which the  orientations of all
particles have a positive projection into one chosen direction is $2^{-N_{cl}}$, which is very small for
clusters of size $10$ or larger. However, there is a sorting mechanism which strongly enhances this
probability, which is the active motion of particles toward the colony edge, where they arrive with roughly
the right orientation \cite{Elgeti_EPL2013}. However, due to the diffusive motion in the bulk of the colony,
see Fig.~\ref{fig:msd}, this mechanism can only operate very close to the colony edge. The polarization profile
(Fig.~3(A) of the main text) shows a high polarization of $p=0.65$ at the boundary. Thus, properly oriented particles 
only have to diffuse laterally along the boundary to form clusters for detachment. This mechanism is 
supported by simulations, which allow the tracking of the history of cluster development.

\subsection{Stress Calculation in a Quasi-One-Dimensional Geometry}

As a further test to our stress estimates, we use force balance to obtain an independent measure of stress, 
similar to traction-force microscopy setups \cite{trepat-nat}. Under the physical interpretation of our system 
that the particles are cells which exert an active force $\gamma v_0$ on the substrate in order to move against 
friction forces $-\gamma v$, the traction force of each particle is $T=\gamma v_0\bf \hat{n}-\gamma \bf v$. 
In one dimension, force balance is closed, and we can calculate a change of stress via force balance. 
We simulate a quasi-one-dimensional geometry with a nearly flat interface. The system consists of a 2D channel 
of dimensions $L_x=6*L_y$ and $L_y=20\sigma$, filled with $N=1200$ particles arranged initially to fill half the channel. 
This system is subjected to periodic boundary conditions in both $x$ and $y$ directions. The stress within 
the cell colony is calculated via integration of force balance (assuming zero stress outside the colony). 

\begin{figure}[ht] 
	\centering
	\includegraphics[width=0.35\textwidth]{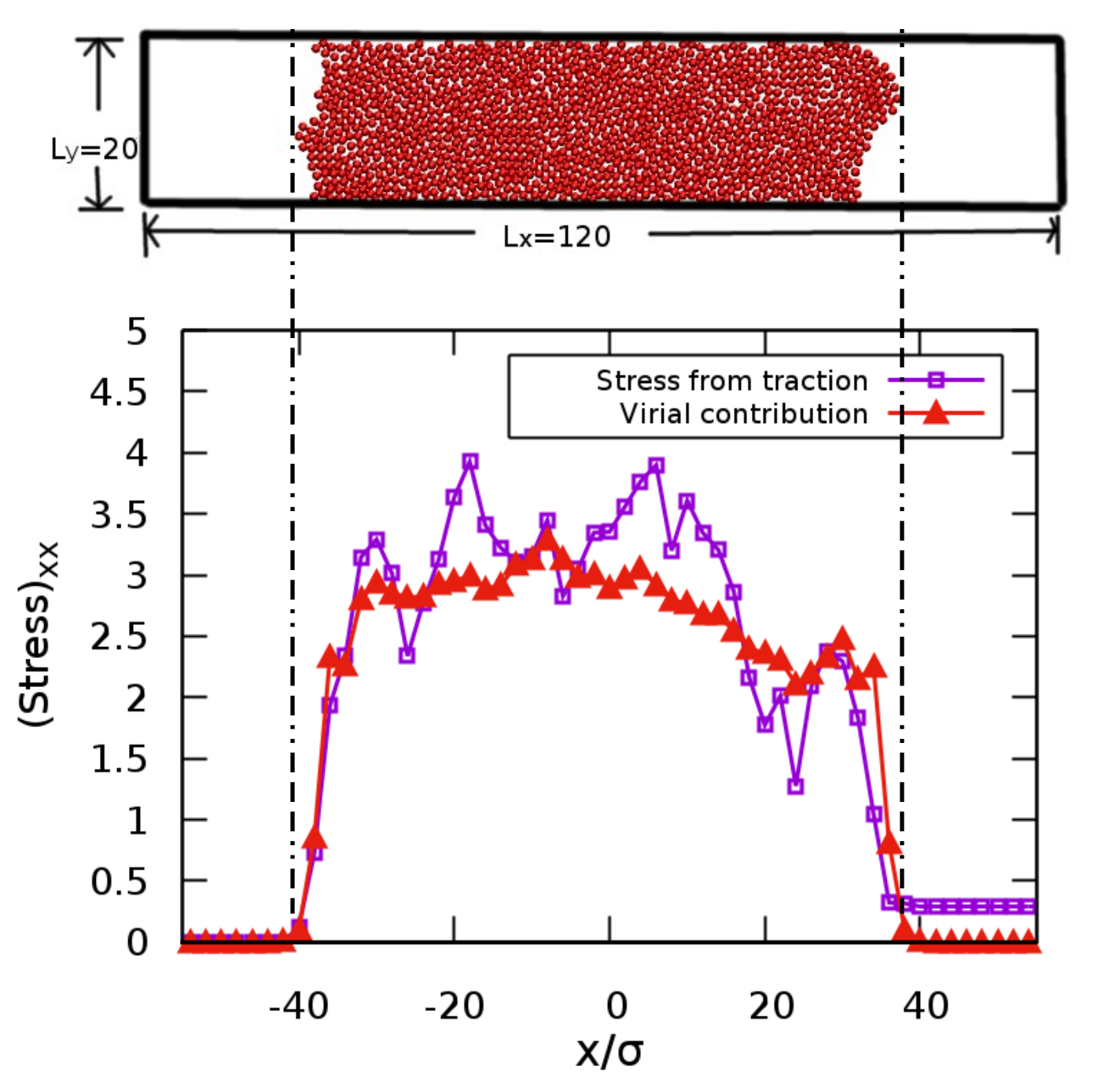}    
	\caption{Top: Snapshot from the simulation in a quasi-one-dimensional rectangular channel, for
		a cell colony in liquid-vacuum coexistence.
		The parameters are $\bar{\sigma}=0.3\sigma$, adhesive strength $U=15$ and activity $Pe=13$.  
		Bottom: Stress calculated from traction forces and virial contribution. }
	\label{fig:strs1} 
\end{figure}

With the parameters $\bar{\sigma}=0.3\sigma$, adhesive strength $U=15$ and activity $Pe=13$, the cell colony is 
in liquid-vacuum coexistence, see Fig.~\ref{fig:strs1}(top). The stress profile in the direction ($x$) normal to 
the interface is shown in Fig.~\ref{fig:strs1}(bottom), while the tangential stress vanishes. 
Figure~\ref{fig:strs1} also shows that the estimations of the stress profile calculated from the traction forces 
and from the virial expression agree quite well. 

\subsection{Origin of Tension in Attractive ABP Clusters --  a Toy Model}

\begin{figure}
	\centering
	\includegraphics[width=0.83\linewidth]{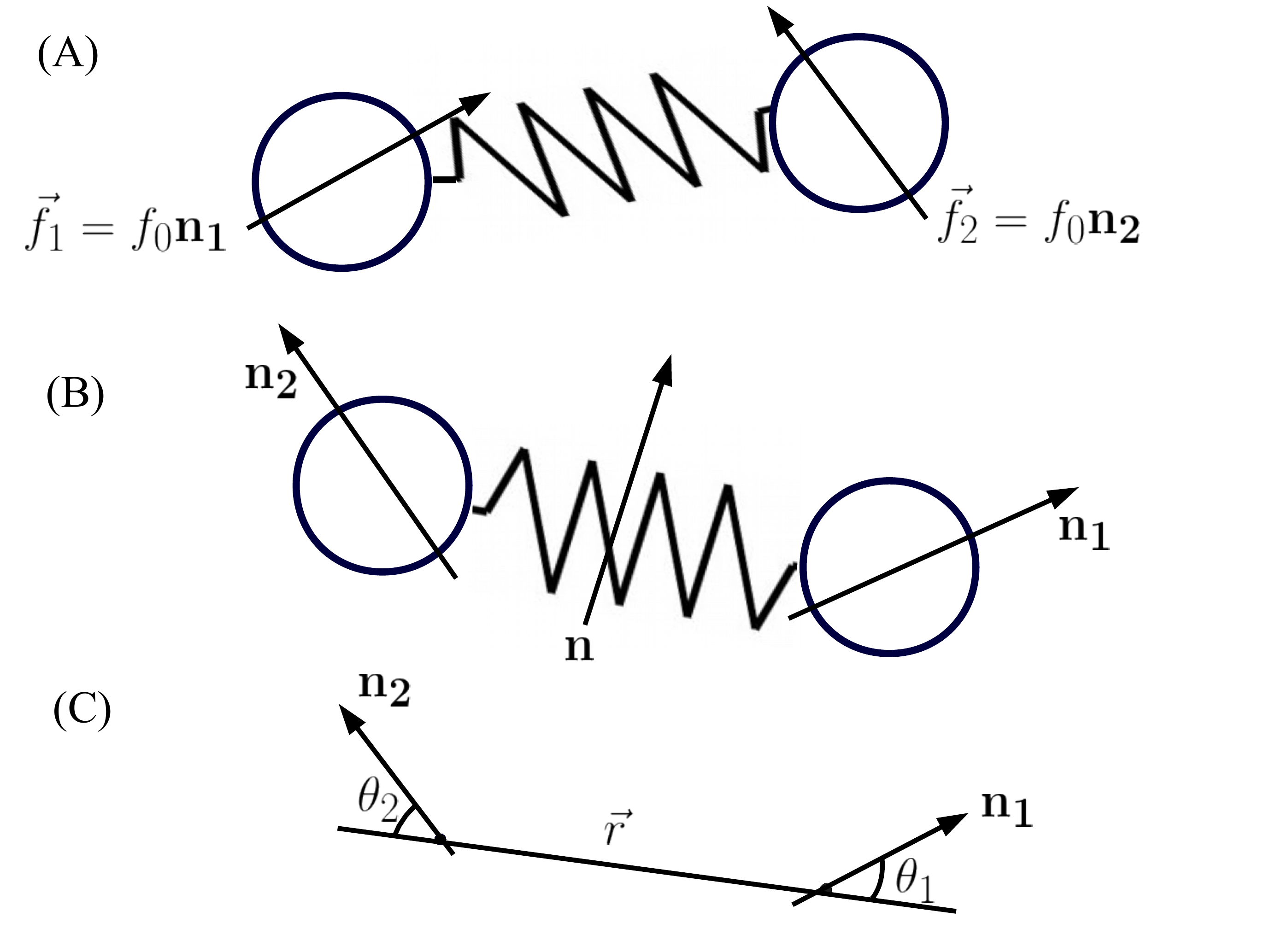}
	\caption{
		Arrangement and forces of ABP dumbbells. 
		(A) Two connected ABPs with randomly oriented propulsion forces ${\bf f}_1$ and ${\bf f}_2$.
		(B) A short time $t$ later, with $\sigma/v_0 < t < \tau_r$, the ABP orientations remain essentially unchanged, but the 
		particles have rearranged to form a quasi-stationary, torque-free state, in which the bond is under tension. 
		The dumbbell also moves in direction ${\bf n}={\bf n}_1 + {\bf n}_2$ normal to the bond vector ${\bf r}$. 
		(C) In the quasi-stationary state, the ABP orientation vectors form angles $\theta_1$ and $\theta_2$ with the bond vector 
		${\bf r} \sim ({\bf n}_1-{\bf n}_2)$, with $\theta_1=\theta_2$. } 
	\label{fig:toy-mdl1}  
\end{figure}

Why are colonies of attractive ABPs under tension, while cluster of repulsive ABPs in the state of
motility-induced phase separation are under pressure? A simple toy model can elucidate the underlying
mechanism. Consider two ABPs which are connected by a harmonic bond \cite{Winkler-2016}. This bond represents 
the interaction between an ABP at the colony edge, and one (or more) ABP further inside the bulk.
In case of a sufficiently high P{\'e}clet number (large propulsion, slow rotational diffusion), the dumbbell 
quickly reaches a quasi-stationary, torque-free state, see Fig.~\ref{fig:toy-mdl1}(A,B). In this state, 
the force can be separated into a propelling component  with direction $({\bf n}_1+{\bf n}_2)$,
normal to the instantaneous bond vector ${\bf r} \sim ({\bf n}_1-{\bf n}_2)$, and a bond stretching
component, see Fig.~\ref{fig:toy-mdl1}(B,C). The stretching force is
\begin{equation}
\begin{aligned}
f_{ext} = & f_0 |{\bf n}_1\cdot \hat{\bf{r}}| = f_0 | \cos(\theta_1) | \\
= & f_0 \sqrt{(1-{\bf n}_1\cdot {\bf n}_2)/2} \\
 = & | \sin((\varphi_1-\varphi_2)/2) |
\end{aligned}
\end{equation}
where $\varphi_i$ is the orientation angle of ${\bf n}_i$ with respect to some fixed axis.
This stretching force has to be averaged over all orientations ${\bf n}_1$ and ${\bf n}_2$, which
yields
\begin{equation}
\langle f_{ext} \rangle = \frac{2}{\pi} f_0 
\label{eq:dumbbell}
\end{equation}
Stress $\Sigma_{\alpha\alpha}$ is force per length, i.e. 
$\Sigma_{\alpha\alpha} \simeq \langle f_{ext} \rangle/a \simeq \langle f_{ext} \rangle/\sigma$.
We can thus use Eq.~\ref{eq:dumbbell} to estimate the tensile stress as a function of P{\'e}clet number. 
With $f_0=\gamma v_0$, $D=k_BT/\gamma$, $D_r=3D/\sigma^2$, and $Pe=3 v_0/(\sigma D_r)$, we obtain $f_0 = Pe$ 
(in our dimensionless units).  Thus, we predict $\Sigma = \Sigma_0 + (2/\pi) Pe$ from our toy-model calculation. 
The linear dependence agrees well with the simulation results for the cell colony, see Fig.~4 of main text. 
However, the slope estimated from Fig.~4 is $0.16$, about a factor 4 smaller than the toy-model
prediction. Two obvious reasons for this overestimation of the slope in the toy model are that (i) the bond
vector takes all orientations with equal probability, but only orientations roughly perpendicular to the
interface contribute to the stress (factor 2), and (ii) the hard-core repulsion between ABPs is neglected,
which sometimes leads to a pressure (negative tension) (maybe another factor 2) -- so that the overall 
agreement is quite satisfactory.

\subsection{Velocity Correlation Function}

\begin{figure}
	\centering
	\includegraphics[width=0.48\linewidth]{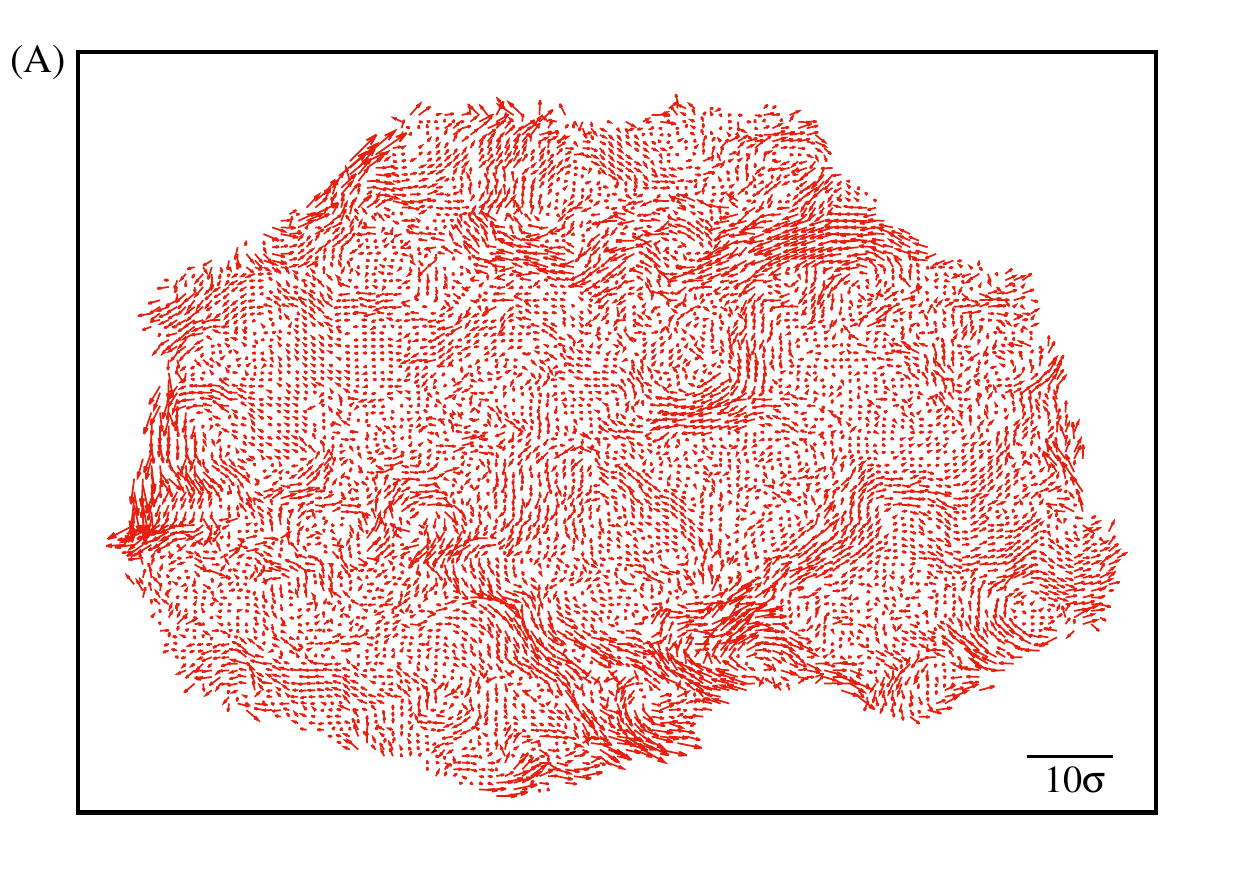}
	\includegraphics[width=0.48\linewidth]{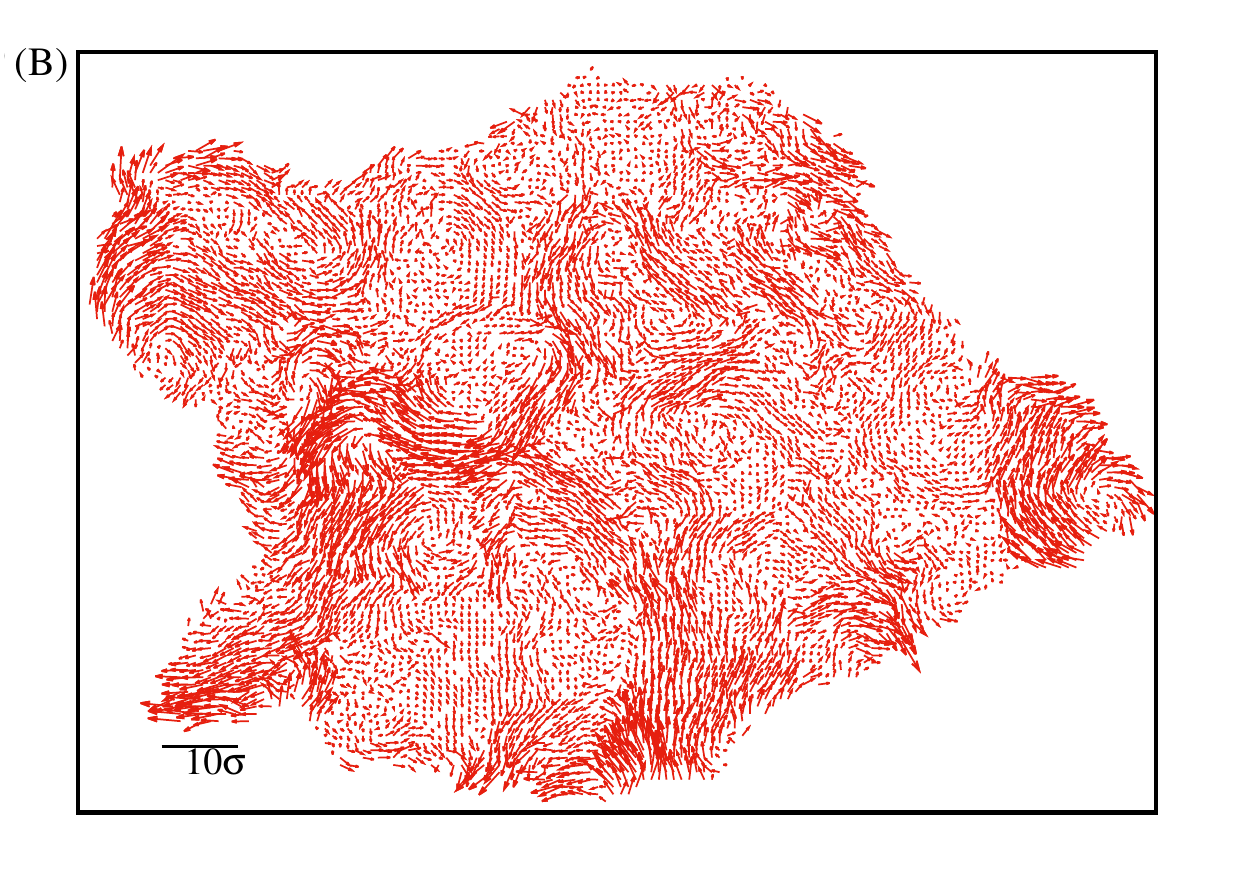}
	\includegraphics[width=0.48\linewidth]{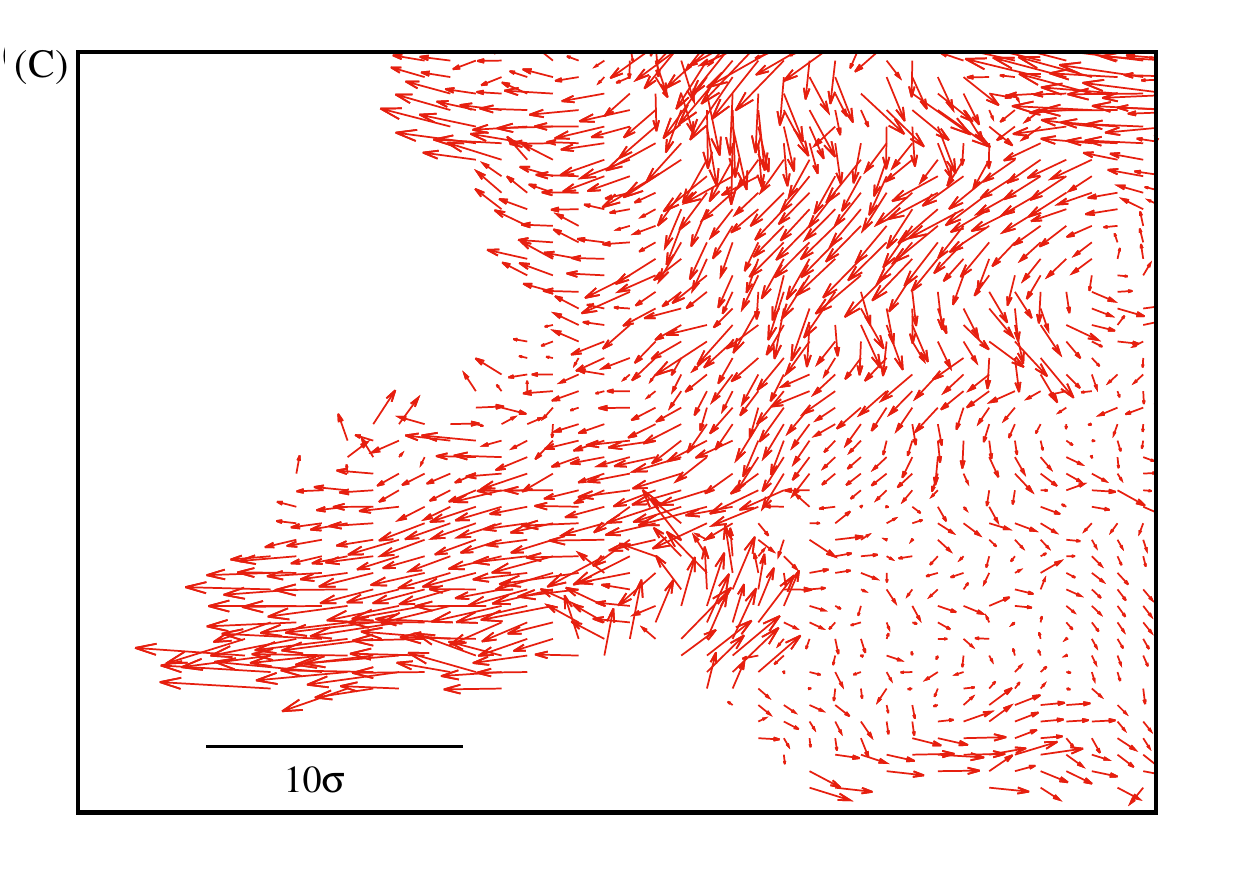}
	\caption{(Top) Velocity field (left) without and (right) with explicit alignment interaction at $U=40$, $Pe=25$ 
		$\bar{\sigma}=0.3$, and $k_e v_0=1.25$. 
		(Bottom) Velocity field in the finger-like structure of the fluid-like  in presence of 
		alignment interaction ($k_e v_0=1.25$).} 
	\label{fig:corltn_snaps}    
\end{figure}

\begin{figure}
	\centering
	\includegraphics[width=0.49\linewidth]{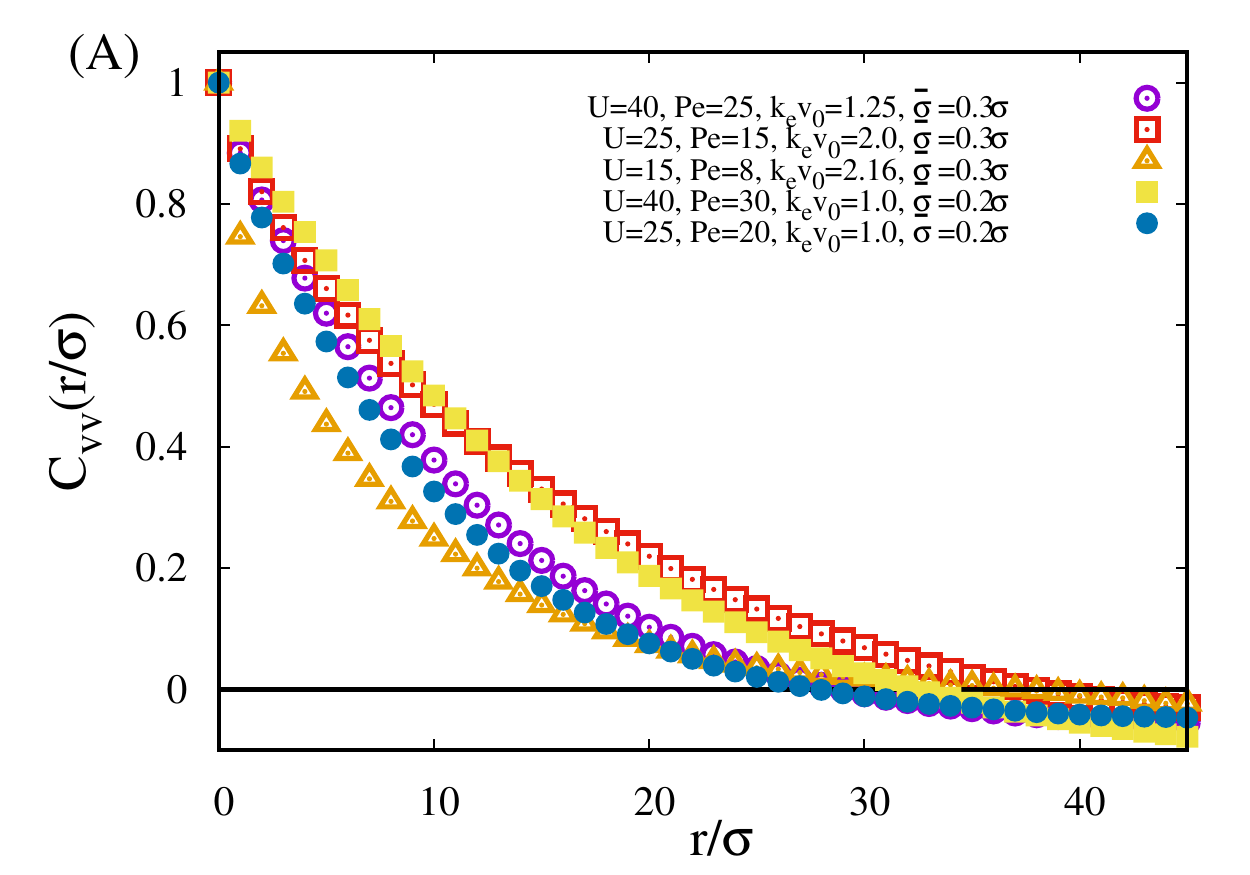}
	\includegraphics[width=0.49\linewidth]{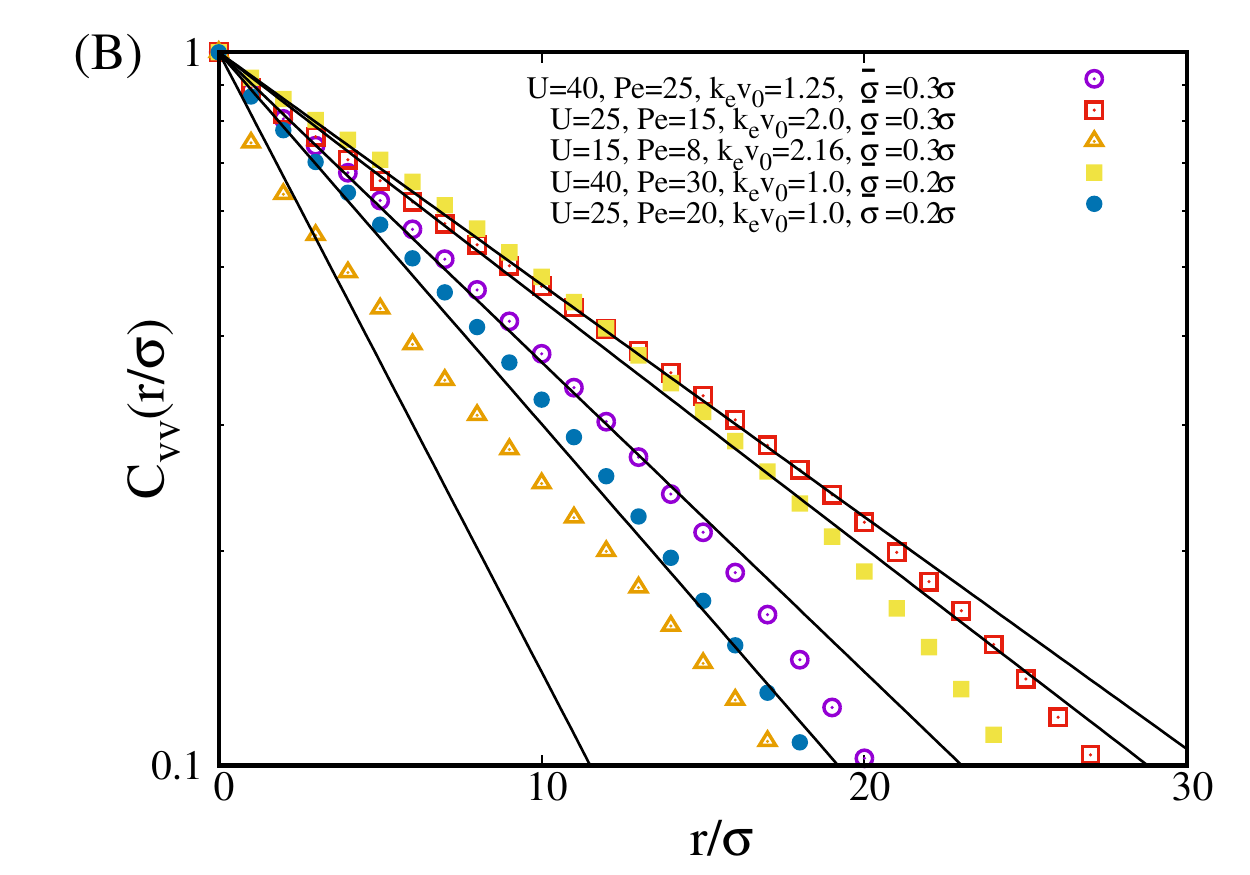}
	\includegraphics[width=0.49\linewidth]{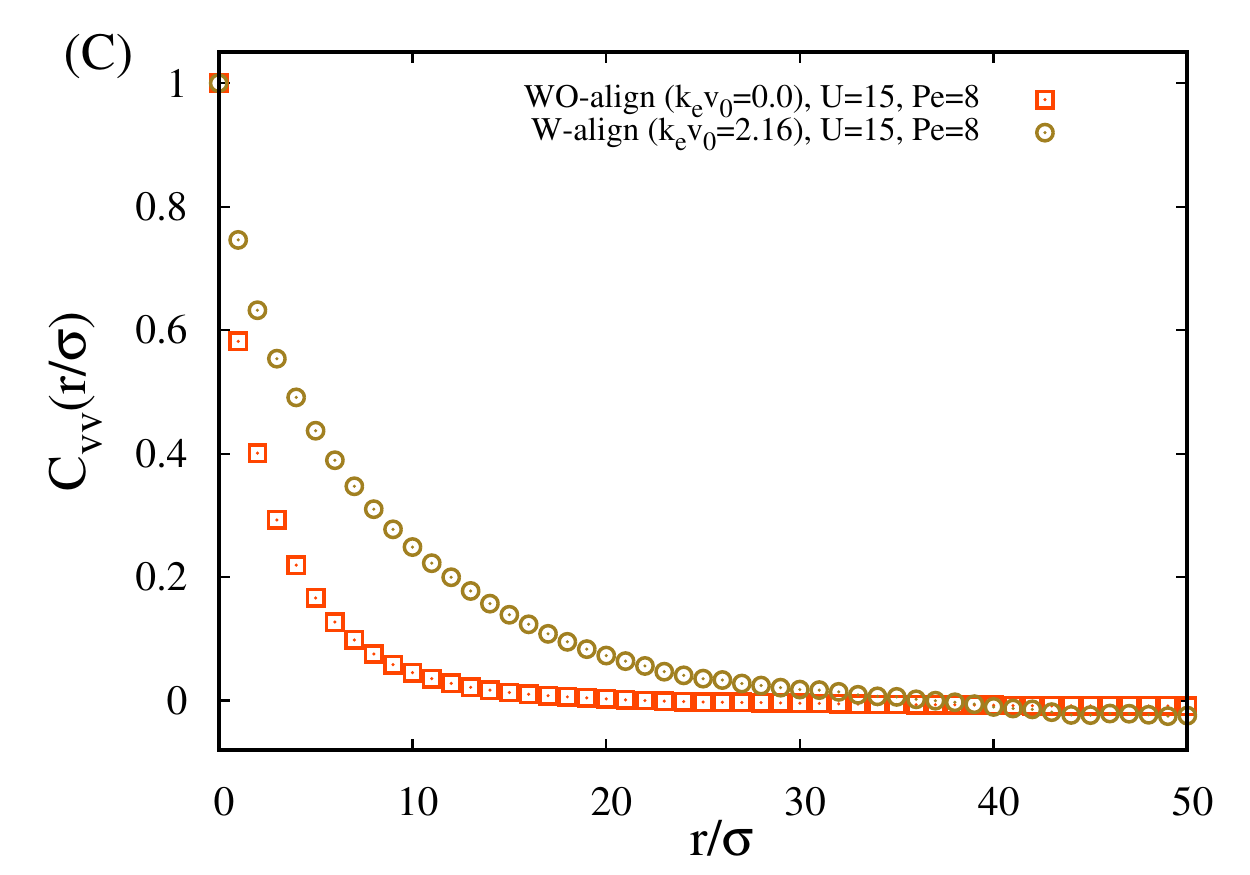}
	\caption{Velocity correlation function $C_{VV}(r/\sigma)$ for a system 
		with velocity-alignment interaction in the L-V region. 
		(A) Spatial dependence for various attraction strengths $U$, P{\'e}clet numbers $Pe$, and alignment
		interaction strengths $k_ev_0$, and widths $\bar\sigma$ of the attraction well, as indicated. 
		(B) Same data as in (A), with exponential decay demonstrated by log-lin representation. 
		(C) Comparison of correlation functions with (w-align) and without velocity alignment (wo-align).}
	\label{fig:corltn}   
\end{figure}

To quantify collective cell migration, we map out the velocity field. Snapshots of the simulations in 
Fig.~\ref{fig:corltn_snaps} demonstrate that particles display strongly coordinated motion. To further quantify 
the correlations, we calculate the velocity correlation function $C_{vv}$ as described in Eq.~5 of the main text. 
Figure~\ref{fig:corltn} shows $C_{vv}(r)$. On short length scales, the velocity correlations decay 
exponentially with a characteristic length scale $\xi_{vv}$. We estimate  
$\xi_{vv}$ by fitting the simulation data by $\exp(-r/\xi_{vv})$. The dependence of the $\xi_{vv}$ on 
$U$ and $Pe$ is discussed in the main text, see Fig.~5(B).

Figure~\ref{fig:corltn} shows $C_{vv}$  with and without alignment interaction. Velocity-alignment interactions strengthen 
correlated motion and result in more swirl-like patterns, as indicated by a small negative minimum in $C_{vv}$. 

\bibliographystyle{apsrmp4-1} 

%

\end{document}